\documentclass[structabstract]{aa}  

\usepackage{graphicx}
\usepackage{natbib}
\bibpunct{(}{)}{;}{a}{}{,}
\usepackage{txfonts}
\usepackage{lscape}
\usepackage{enumerate}
\usepackage{color}

\newcommand\T{\rule{0pt}{2.6ex}}
\newcommand\B{\rule[-1.2ex]{0pt}{0pt}}
\begin{document}
   \title{Rest-frame stacking of 2XMM catalog sources}
 \subtitle {Properties of the Fe K$\alpha$ line}

 	\author{P. Chaudhary
           \inst{1}
           \and
           M. Brusa\inst{1}
  	  \and	
   	  G. Hasinger\inst{2,3}
   	  \and
   	A. Merloni\inst{1}
  	   \and
  	A. Comastri\inst{4}
	    \and
	K. Nandra\inst{1}
	 }

  \institute{Max-Planck-Institut f\"ur  extraterrestrische Physik,
  Giessenbachstrasse 1, 85748 Garching bei M\"unchen, Germany\\
        \email{chaudhary@mpe.mpg.de}
        \and
	Max-Planck-Institut f\"ur Plasmaphysik, Boltzmannstrasse 2, 85748 Garching bei M\"unchen, Germany	
	\and
	Institute for Astronomy, 2680 Woodlawn Drive, Honolulu, HI 96822-1839, USA
	\and
	 INAF - Osservatorio Astronomico di Bologna, via Ranzani 1, 40127 Bologna, Italy \\
              }
\date{Received ; accepted}

\abstract{}{}{}{}{} 

\abstract
{}
{The aim of this work is to characterize the average Fe K emission properties of active galactic nuclei (AGNs) in the source rest-frame.}
{We selected a sample of 248 AGNs from the 2XMM catalog, covering a wide redshift range $0 < z < 5$ and with the EPIC-PN net 2--10 keV rest-frame counts $\geq$200 and power law photon indices in the range 1.5--2.2. We employed two fully independent rest-frame stacking procedures to compute the mean Fe K profile. The counting statistics for the integrated spectrum is comparable to the one available for the best studied local Seyferts. To identify the artifacts possibly introduced by the stacking procedure, we have carried out simulations.}
{We report that the average Fe K line profile in our sample is best represented by a combination of a narrow and a broad line. The equivalent widths of the narrow and broad (parametrized with a {\tt diskline}) components are $\sim$30 eV and $\sim$100 eV, respectively. We also discuss the results of more complex fits and the implications of the adopted continuum modeling on the broad line parameters and its detection significance.}
{}

\maketitle
%
\section{Introduction}
  
Fluorescent Fe K$\alpha$ lines in active galactic nuclei (AGNs) are a potentially unique tool to probe the innermost regions around a black hole. The line properties, such as the peak energy, intensity and profile carry important diagnostic information about the dynamics and physics of the region where the emission originates. The measured line energy can be used to infer the ionization state of the line emitting matter, whereas the line equivalent width (EW) indicates the amount of fluorescing material \citep{Fabian2000}. 

The line profile is determined by several physical parameters. The Fe K line in AGNs is assumed to be produced through X-ray irradiation of optically thick matter, such as the molecular torus and/or the accretion disk \citep{Guilbert1988, Lightman1988, George1991, Matt1991}. If the Fe K$\alpha$ line arises in distant material like the molecular torus envisaged in orientation--dependent unification scheme for AGNs \citep{Antonucci1993}, the resulting emission line profile is narrow. In contrast, the Fe K$\alpha$ line originating in the inner accretion disk is distorted by Doppler and gravitational effects and becomes asymmetric \citep{Fabian1989, Laor1991}. Hence, relativistic lines offer a robust way to measure accretion disk properties, such as the radial extent, the emissivity profile, the inclination angle of the disk to the observer's line of sight and the black hole spin \citep[for the reviews see][]{Reynolds2003, Fabian2005, Miller2007, Turner2009}.  

The last decade has witnessed a significant improvement in our knowledge of the narrow Fe K$\alpha$ line properties in AGNs. Based on data collected with high-sensitivity X-ray satellites XMM-{\it Newton}, {\it Chandra} and {\it Suzaku} the ubiquitous presence of a neutral, narrow Fe K$\alpha$ line in AGNs spectra has been established \citep{Yaqoob2004, Gu2006, Reeves2006, Nandra2007}. The behavior of the Fe K$\alpha$ line as a function of the X-luminosity and redshift has also been examined. Compelling evidence of the inverse correlation between the neutral, narrow Fe K$\alpha$ line EW and X-ray luminosity, known as the X-ray Baldwin effect (Iwasawa-Taniguci effect), has been reported \citep{IT1993, Nandra1997, Page2004, Bianchi2007, Chaudhary2010, Shu2010}. However, the EW is observed to be independent of the redshift \citep{Brusa2005, Chaudhary2010}. 

Since the first unambiguous detection of relativistically broadened Fe K$\alpha$ emission in the X-ray spectrum of MCG-6-30-15 observed with the ASCA satellite \citep{Tanaka1995}, sincere efforts have been made to test for the presence of broad Fe K$\alpha$ emission lines in AGNs. Recent X-ray observatories like {\it Chandra}, XMM-{\it Newton} and {\it Suzaku} have revealed that statistically significant broad Fe lines can be detected in relatively deep observations. Using a sample of 102 AGNs observed with XMM-{\it Newton} at $z < 0.5$, \citet{Gu2006} detected relativistic lines in about 25\% of the sample objects ($\simeq$50\% when the ``well-exposed'' spectra with $\gtrsim$10 000 net counts in the 2--10 keV band are considered). They also stacked residuals in four equally populated luminosity classes and find that the average EWs are $\lesssim$150 eV for all luminosity classes. \citet{Nandra2007} performed a spectral analysis of a sample of 26 nearby Seyfert galaxies ($z < 0.05$) observed by XMM-{\it Newton} with a minimum of 30 000 net counts in the EPIC-PN spectrum in the 2--10 keV band. They found evidence of broad line emission in around 65\% of the sample with a typical EW of $\sim$77 $\pm$ 16 eV, when fitted with a broad Gaussian. \citet{Dela2010} carried out a systematic and uniform analysis of 149 local radio quiet Type 1 AGNs and found strong evidence of a relativistic Fe K$\alpha$ line in 36\% of the sources, with an average EW of 100 eV \citep[see also][]{Patrick2010}. 
 
The X-ray spectra of the distant AGNs are often photon starved. The low photon counting statistics in such AGNs spectra prevents an appropriate modeling of the underlying continuum and therefore the significance of the broad line component and its measured parameters suffer from considerable uncertainties. The stacking analysis, a powerful technique to improve the counting statistics, can be applied to gain an insight into the average properties of faint AGNs. 

Using a large sample of 352 AGNs detected in the Chandra Deep Field North and South, \citet{Brusa2005} computed the average spectra 
by stacking X-ray counts in the observed frame in seven redshift bins over the range $0.5 < z < 4$. In a previous paper \citep[][ hereafter CH10]{Chaudhary2010}, we assembled a sample of 507 sources from the 2XMM catalog at high galactic latitude ($|BII|$ $>$ 25 degrees), with the sum of the EPIC-PN and EPIC-MOS 0.2--12 keV counts greater than 1000 and covering the redshift range $0 < z < 5$. We stacked X-ray spectra in the observed frame in narrow redshift bins and investigated the fundamental properties of the integrated spectrum, such as the ubiquity of the Fe K$\alpha$ emission in AGNs and the dependence of the spectral parameters on the X-ray luminosity and redshift. However, the analyses of \citet{Brusa2005} and CH10 were limited to the narrow line properties as the broad line parametrization was hampered by redshift smearing in the observed frame stacked spectra.    

\citet{Alina2005} constructed an average rest-frame spectrum of 104 X-ray sources (53 Type 1 and 41 Type 2 AGNs and 10 galaxies) covering $0 < z < 4.5$ in redshift space and reported a clear relativistic line in the average rest-frame spectra with an EW of $\sim$560 eV and $\sim$460 eV for the Type 1 and Type 2 AGNs, respectively. \citet{Longinotti2008} stacked X-ray spectra from a local sample of 157 AGNs and found that the EW of the broad relativistic line is never higher than 100 eV, either when stacking the whole sample or different sub-samples. \citet{Corral2008} computed the mean Fe emission from a large sample of more than 600 Type 1 AGNs spanning a redshift range up to $\sim$3.5. They detected significant narrow Fe K$\alpha$ emission line around 6.4 keV with an EW$\sim$90 eV and found no compelling evidence of any significant broad relativistic emission line in the average spectrum \citep[see also][]{Mao2010}. 

While the presence of a narrow line component is confirmed by all independent analyses, the evidence of the presence of a broad line is more debated. The complex nature of AGNs spectra, and, in particular, the degeneracy between the spectral parameters and the dependence of the line EW and shape over the fitted (absorbed) continuum, hampered so far a comprehensive interpretation of the physics behind the observed features. As an example, it has also been proposed that the observed red-wing can also be ascribed to complex (clumpy) absorption at least for MCG--6-30-15 \citep{Miller2009}. It is therefore critical to have multiple independent procedures to determine the Fe line parameters. 

In this paper, we study a small but well-defined sample of AGNs taken from CH10, and concentrate on the mean properties of the Fe K line in the rest-frame using two fully independent procedures. The paper is organized as follows: Section \ref{sect:Sample} describes the selection criteria and properties of the sample. Sect. \ref{sect:RF-stacking-proc} outlines the stacking procedures. The results of the spectral analysis are presented in Sect. \ref{sect:Results} and are discussed in Sect. \ref{sect:Discussion}. Throughout this work a cosmology with $\Omega_m = 0.27$, $\Omega_\Lambda = 0.73$ and $H_0 = 70 $ km s$^{-1}$ Mpc$^{-1}$ is used. Spectral analysis was carried out with {\tt XSPEC} (version 12.6.0); errors are reported at the 68$\%$ confidence level for one interesting parameter ($\Delta\chi^2 = 1.0$). All energies refer to the rest-frame unless otherwise specified.

\section{The sample}\label{sect:Sample}
We start with the CH10 sample of 507 AGNs at high galactic latitude ($|BII|$ $>$ 25 degrees), with the sum of the EPIC-PN and EPIC-MOS 0.2--12 keV counts greater than 1000 and covering the redshift range $0 < z < 5$. This sample was selected from the 2XMM catalog, the second comprehensive catalog of serendipitous X-ray sources from the European Space Agency's (ESA) XMM-{\it Newton} observatory. The catalog contains $\sim$250 000 X-ray source detections which relate to $\sim$200 000 unique X-ray sources \citep{Watson2009}. Redshifts were obtained from the Nasa's Extragalactic Database. For each object, source specific products such as the source and background spectra, images in the different energy bands and response files were retrieved from the XMM-{\it Newton} Science Archive. We refer the reader to CH10 for all the details on the selection of the sample and safety checks performed to assess the quality of the archival products. 
  
\begin{figure*}
  \begin{center}
    \begin{tabular}{cc}
      \includegraphics[height=6.0cm]{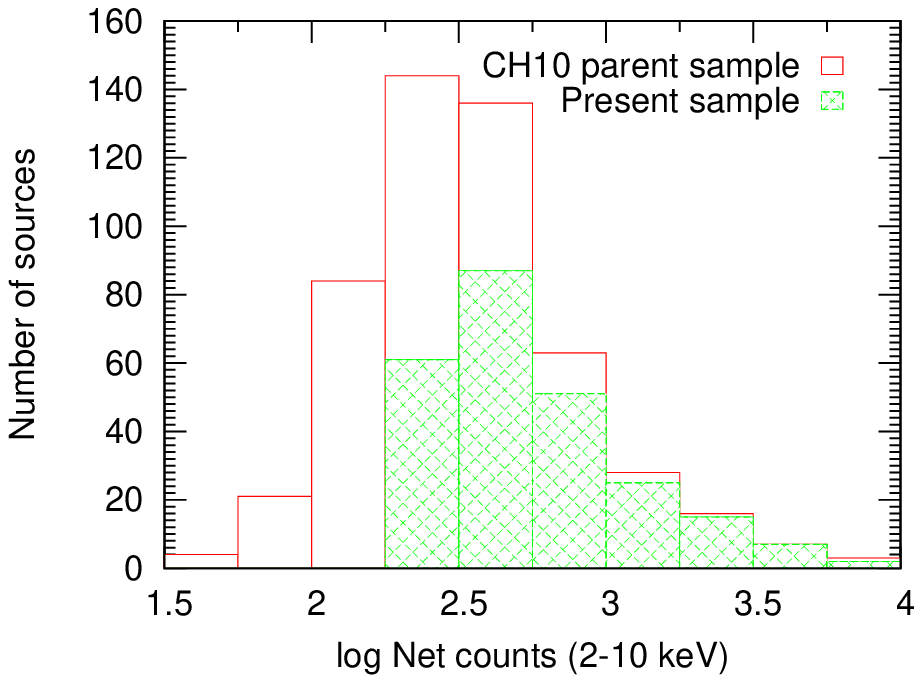} &
      \includegraphics[height=6.0cm]{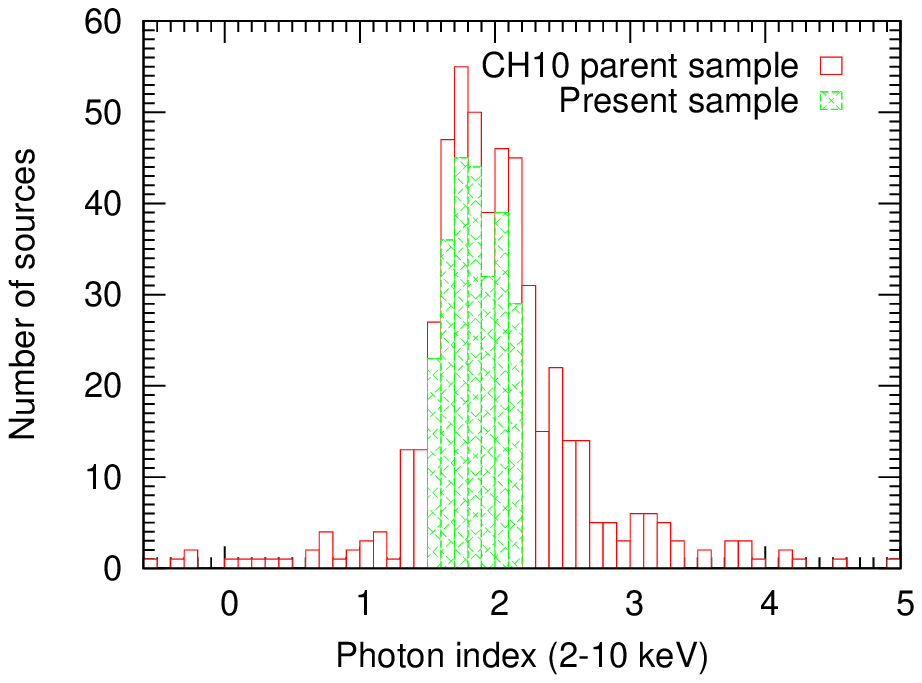}\\
      \includegraphics[height=6.0cm]{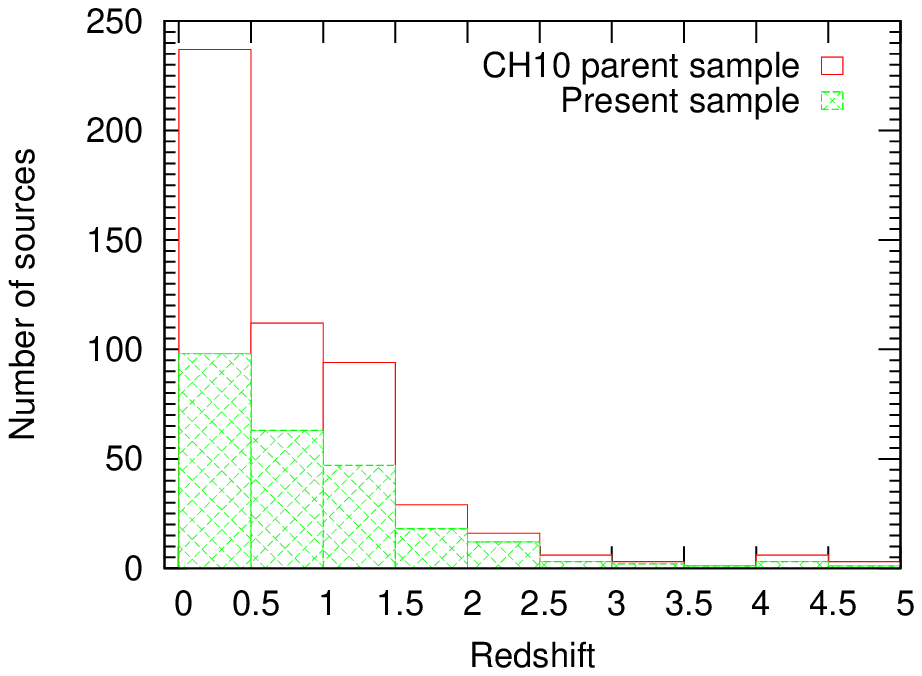} &
      \includegraphics[height=6.0cm]{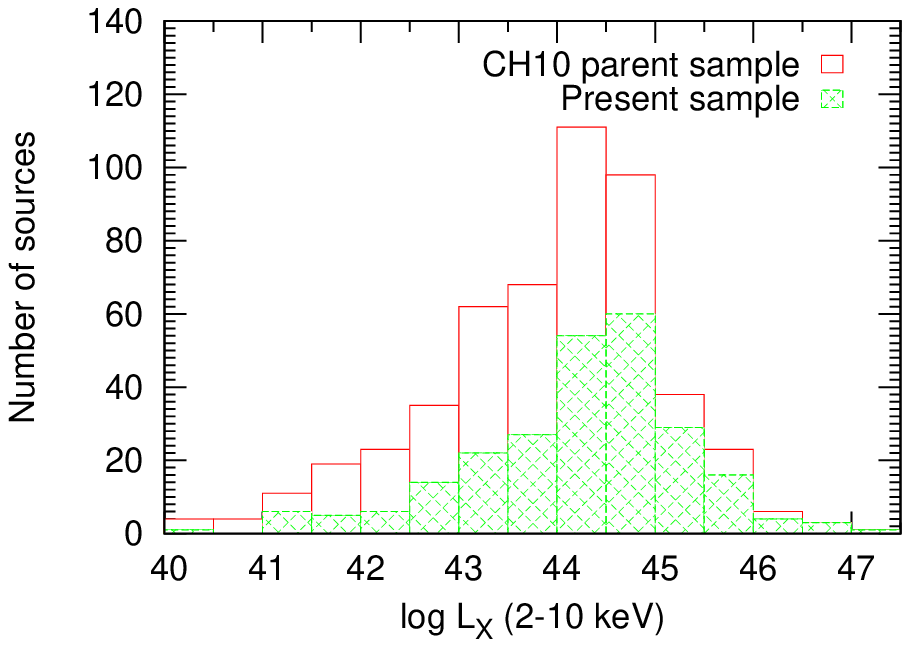} \\
      
    \end{tabular}
  \end{center}
  \caption{\textit{Top:} distributions of the EPIC-PN net counts (\textit{left}) and 2--10 keV power law photon indices (\textit{right}) of the 248 sources selected for our study (green cross-hatched) compared with the CH10 parent sample (red solid line). \textit{Bottom:} redshift (\textit{left}) and rest-frame 2--10 keV luminosity, no absorption correction (\textit{right}) distributions of these two samples.}
\label{fig:subsample}
\end{figure*}

An essential requirement of the conventional rest-frame stacking analysis is the spectrum unfolding by appropriately modeling the underlying continuum. Therefore, to carry out a reliable spectral analysis of each spectrum, we refined the sample of CH10 by considering only the AGNs with the EPIC-PN net 2--10 keV rest-frame counts $\geq$200 ($\sim$73\% of the parent CH10 sample). As our main interest is in the Fe K region of the spectrum, we performed a detailed spectral analysis of the individual spectrum in the rest-frame 2--10 keV energy range, excluding the 5.5-7.0 keV Fe band\footnote{Spectral fits excluding 5--7 keV and 6--7 keV have also been performed. This did not change the results.} using {\tt XSPEC} \citep[ver. 12.6.0;][]{Arnaud1996}. We fitted the data with an absorbed power law\footnote{We also tested a model with a neutral reflection component \citep[{\tt pexrav} in {\tt XSPEC}, see][]{Mag1995}. However, our choice of the spectral energy range (2--10 keV) and the limited  statistics in the majority of spectra (see Fig. \ref{fig:subsample} for sample properties) restrained a full parametrization of the reflected continuum. We therefore adopted a simple absorbed power law continuum for all sources.}. Only the EPIC-PN data has been used in this work.

From the original CH10 sample of 507 sources, we further selected a sample of 248 AGNs having their power law photon indices as derived from the above mentioned fits in the range 1.5 $\leq \Gamma \leq$ 2.2. The imposed threshold on $\Gamma$ ensured that our results are stable against the continuum normalization and the contribution of the highly obscured AGNs with prominent Fe lines is minimum. The top row in Fig. \ref{fig:subsample} shows the distributions of the EPIC-PN net counts (left) and power law photon indices in the rest-frame 2--10 keV band (right) of the 248 sources used in this work (green cross-hatched) compared to the CH10 parent sample (red solid line). The bottom row shows overlaid histograms of redshifts (left) and 2--10 keV luminosities, not corrected for absorption (right) for these two samples. We note that the 248 sources subsample can be considered a fair representative of the CH10 sample in terms of the X-ray luminosity and redshift.    

\section{Rest-frame stacking procedures}\label{sect:RF-stacking-proc}
\subsection{Stacked ratio}\label{sect:RF-stacked-ratio-proc}
A commonly employed method to study the average Fe line properties is to derive the average of the ratio between the data and a simple continuum model for each source under study \citep[e.g. ] []{Nandra1997, Gu2006}. However, several key factors should be considered in the stacking analysis when the sample sources have a wide redshift distribution. These include different redshift of each source and thus a different observed line energy, the energy dependent detector response is also different for each source.  

Taking into account all these factors we adopted the following procedure for constructing an average ratio profile for our AGNs sample. 

\begin{figure}
 \centering
\includegraphics[height=6cm]{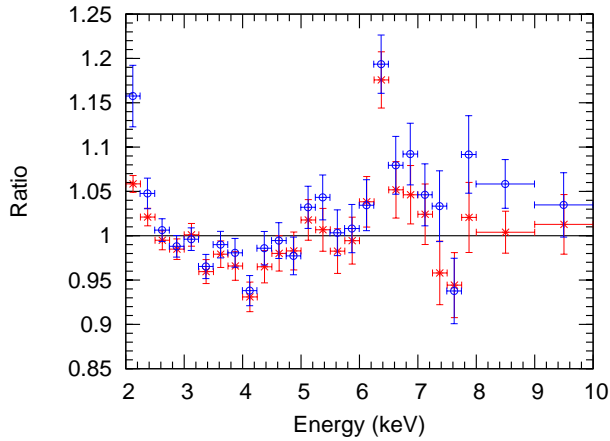}
\caption{Comparison of the mean ratio profiles: the whole sample comprising 248 sources (blue) and after applying 3--sigma clipping on the ratios in each energy bin (red).}
\label{fig:avgratio}
\end{figure}

\begin{itemize}
\item[$\bullet$] For each source, we determined the observed frame energy range corresponding to the rest-frame 2--10 keV band and the associated ungrouped channel information from the respective response files. Each spectrum, in the rest-frame 2--10 keV band, was then grouped in predefined bins of widths equal to $0.25/(1 + z)$ up to 8 keV and $1/(1 + z)$ in the 8--10 keV using the {\tt FTOOLS} routine {\tt grppha}, which corresponds to a rest-frame sampling of 0.25 keV and 1 keV in the 2--8 keV and 8--10 keV, respectively. Thus, we had 26 energy bins in each spectrum.
\item[$\bullet$] Given the limiting counting statistics and the choice of predefined binning we could not assure a minimum of 20 counts per bin. We therefore analyzed the binned, background-subtracted, 2--10 keV rest-frame spectra with an absorbed power law using the Cash statistic \citep{Cash1979} implemented in {\tt XSPEC}. The best fit parameters (N$_H$ and $\Gamma$ etc. with errors) and ratio with respect to the best fit continuum model were saved. 

\item[$\bullet$] These ratios were then summed and averaged for all the sources in each rest-frame energy bin. First, we derived the mean and standard deviation of the ratios in each energy bin. We then removed in each energy bin all the ratios deviating more than 3 times the standard deviation from the average value (3--sigma clipping). Figure \ref{fig:avgratio} shows the resulting averaged ratio (red) comprising $\geq$95\% of ratios in each energy bin compared with the mean ratio for the whole sample of the 248 sources (blue). We find that the two ratio profiles are consistent within the errors as expected. We note that the 3--sigma clipped averaged ratio (red) is comparatively lower at high energies ($>$ 7 keV) as the contribution of the ratios deviating significantly from the average value in these bins is removed.  
  
\item[$\bullet$] The averaged ratio cannot be fitted directly in {\tt XSPEC} and it must be converted to a flux spectrum taking into account the average underlying continuum, we therefore converted the averaged ratio created by applying 3--sigma clipping (red points in Fig.  \ref{fig:avgratio}) to a flux spectrum by multiplying the averaged ratio in each energy bin by the E$^{-\Gamma}$ factor, where E is the central rest-frame energy of the bin and $\Gamma$(=1.8) is the mean slope of our sample. We refer to this flux spectrum as the ``averaged ratio flux spectrum''.
\end{itemize}

\subsection{Stacked spectrum}
We followed the averaging procedure of Iwasawa et al. (2011) for constructing an ``averaged X-ray spectrum''. The advantage of this procedure is that it is independent of the continuum modeling and statistics used in the spectral analysis. In this procedure original data, in units of counts s$^{-1}$ keV$^{-1}$, of each spectrum were saved after loading the spectrum grouped as described above. We also created the ascii tables of the ancillary response matrices (.arf) with the same binning as of the spectrum. The data were then divided by the average effective area in each energy bin, and thereafter normalized to the 3--5 keV continuum to ensure that the average spectrum is not dominated by the brightest objects. The final ``averaged X-ray spectrum'' is constructed by summing the normalized counts of all the different sources in each rest-frame energy bin and averaging for the total number of sources. A special effort is made to evaluate calibration issues and uncertainties (see Iwasawa et al. (2011) for further details). The total number of the 2--10 keV counts for the 248 sources is $\sim$198000. 

\begin{figure}
\centering
\includegraphics[height=6cm]{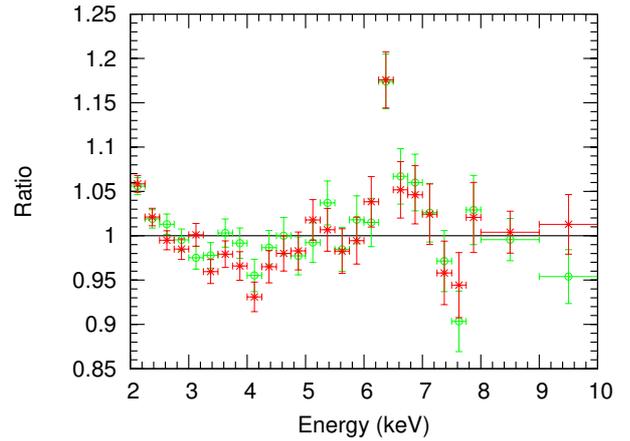} 
\caption{Ratio of the ``averaged X-ray spectrum'' with respect to a power law (green) overlaid with the mean ratio profile of the 248 sources created using 3--sigma clipping (red). }
 \label{fig:avgratio-from-avgspec}
 \end{figure}

The ratio of the ``averaged X-ray spectrum'' with respect to a power law (green) over-plotted with the mean ratio profile created using 3--sigma clipping (red) is shown in Fig. \ref{fig:avgratio-from-avgspec}. We notice that the two ratio profiles are fully consistent within the statistical errors. Spectral fitting of the ``averaged X-ray spectrum'' is addressed in more detail in Sect. \ref{sect:Avgspec-fitting}.
     
\subsection{Simulations}\label{sect:Simulations} 
\begin{figure}
\centering
\includegraphics[height=6cm]{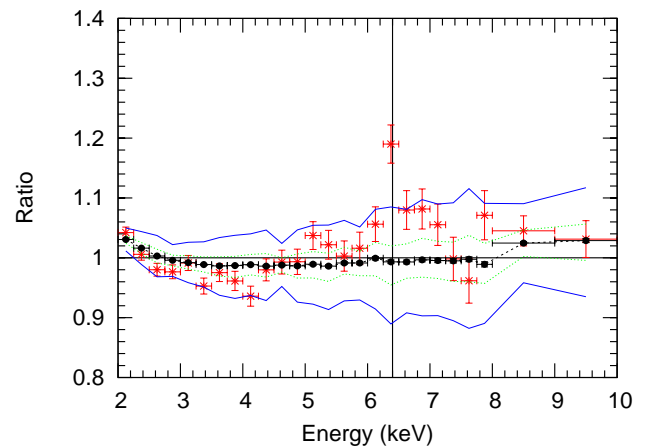} 
\caption{Averaged ratio of the 248 spectra after applying 3--sigma clipping (red), mean simulated continuum (black) along with its 1$\sigma$ (green dotted line) and 3$\sigma$ (blue solid line) confidence limits. The solid vertical line is drawn at 6.4 keV.}
\label{fig:simulations-cf}
\end{figure}

To identify the artifacts possibly introduced by the stacking method, we carried out simulations using {\tt XSPEC}. First, we created 100 simulated ``ungrouped'' line less spectra for each of the 248 sources using their best spectral fit parameters (e.g, N$_H$, $\Gamma$ and power law normalization etc.). Identical response and effective area files were used for the real and simulated spectra. 

We then applied the same procedure detailed in Sect. \ref{sect:RF-stacked-ratio-proc} to these simulated spectra and computed the mean ratio profile of the 248 simulated spectra for each realization. Thus, we have 100 realizations of the underlying continuum represented by the mean ratio of each realization. We then computed the distribution of these simulated continua. The averaged ratio profile of the 248 spectra (red), mean simulated continuum (black) and its 1$\sigma$ (green dotted line) and 3$\sigma$ (blue solid line) deviations in each energy bin are shown in Fig. \ref{fig:simulations-cf}.  

We note that some apparent deviations (e.g. upward turn at energies below 2.5 keV) present in the averaged ratio profile is also evident in the averaged simulated continuum. This in turn, confirms that simulations can identify very small systematic effects of the order of 5\% and the significance of any spectral feature can be deduced with a reliable accuracy. We have investigated the possible reasons for this curvature at energies below 2.5 keV. This effect can be attributed to a combination of the boundary effect (due to our spectral energy range selection) and counting statistics. Indeed, when the averaged ratio profile and the averaged simulated continuum are computed for the sources with net counts $>$500, we do not find any curvature.  

\section{Results}\label{sect:Results}

\subsection{``Averaged ratio flux spectrum'' fitting}\label{sect:Avgratio-fitting}
We performed a detailed spectral analysis of the ``averaged ratio flux spectrum'' constructed from the averaged ratio profile as described in Sect. \ref{sect:RF-stacked-ratio-proc}. All the fit results are shown in the Appendix. We used the {\tt FTOOLS} routine {\tt flx2xsp} to convert the ascii file of the ``averaged ratio flux spectrum'' to the fits format. To assess the improvement in the spectral fitting on including an additional spectral feature in the fit, we adopt the $\Delta\chi^2$  criterion (e.g. $\Delta\chi^2$ = 2.7 corresponds to 90\% significance for the addition of one interesting parameter and $\Delta\chi^2$ = 4.6 corresponds to 90\% significance for two additional interesting parameters). Throughout the spectral analysis, errors and upper limits are quoted at the 68$\%$ confidence level for one parameter of interest. The errors in the Fe line EW are calculated using {\tt XSPEC} {\it eqw with err option}.

We first carried out the spectral analysis in the 2--10 keV energy range using various models consisting of simple/complex continuum and one/two Fe K$\alpha$ lines. However, in all the attempted fits, we noticed the presence of significant residuals of the order of 5\% in the 2--3 keV band that can be ascribed to the systematic effect identified by our simulations discussed in section \ref{sect:Simulations}. Moreover, the 2--3 keV excess resulted in high $\chi^2$ values (e.g. $\chi^2 > 2$) in most of the models explored. Therefore, to obtain more reliable spectral fits, we restricted the spectral analysis of the averaged ratio flux spectrum in the 3--10 keV band. 
 
We then concentrated on the characterization of the narrow Fe K line. To parametrize the narrow core we used a power law plus a narrow Gaussian line with the energy and width fixed at 6.4 keV and 0.01 keV, respectively. The line normalization was allowed to vary. This line model yielded an acceptable fit ($\chi^2/dof = 31.3/19$) and the narrow line is detected at $>$99.995\% significance level ($\Delta\chi^2 \sim$ 30 for one additional parameter), with an EW of $45 \pm 13$ eV. We note that the residuals show excess emission of the order of a few per cent in the 6--7 keV energy range. 

To search for the presence of a broad component we replaced the narrow feature with a broad Gaussian by leaving the line width free to vary. The fit was considerably better than a single narrow Gaussian emission line model ($\chi^2/dof = 25.6/18$), with resulting values of the broad line EW of $75 \pm 40$ eV and a measured width of $\sigma = 0.15^{+0.06}_{-0.04}$ keV. However, we see excess emission around 6.7 keV possibly indicating that the model with a single line component does not fully describe the data. This hypothesis is further supported by the observed improvement in the goodness of fit as a single Fe K line with free width is partially modeling the broad feature. 

We next fitted the data with a simple power law and a physically motivated broad line with an emission profile from an accretion disk around a Schwarzschild black hole \citep[{\tt diskline},][] {Fabian1989}. We fixed several {\tt diskline} parameters; the peak energy, $E_{Disk}$, at 6.4 keV, the inner disk radius, $R_{in}$, outer disk radius, $R_{out}$, at 6 and 100 $R_g$, respectively, where $R_g$ is the gravitational radius and the emissivity index, $\beta$, at $-2.5$. The free parameters were the disk inclination, $i$, and normalization. The fit was statistically worse ($\chi^2/dof = 41.6/18$) as compared to the previous ones, which can be attributed to the presence of significant residuals peaking at 6.4 keV. The errors in the fit parameters except the {\tt diskline} EW ($128 \pm 41$ eV) could not be inferred due to the poor fit. Thus, we concluded that a line model consisting of two components should be used to model the Fe K band in our data. 

\begin{figure}
\centering
\includegraphics[height=9cm, angle=-90]{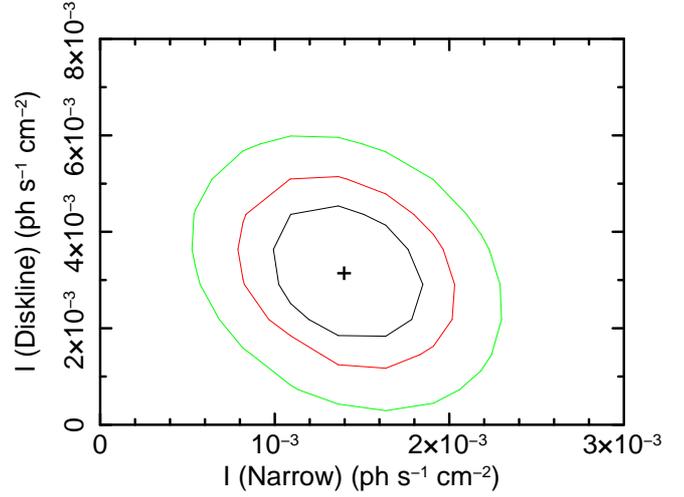}
\caption{Contour plots (68\%, 90\% and 99\%) of the {\tt diskline} versus narrow component intensities for the ``averaged ratio flux spectrum'' . A simple power law is used to model the underlying continuum. The cross indicates the best fit values.}
\label{fig:cont-pow-nc-disk-rflx}
\end{figure}

We therefore tested a two component line model consisting of two Gaussians to account for the narrow core and the broad component. The fixed parameters for the narrow component were the same as those in the single narrow Gaussian line fit mentioned above. The line energy of the broad feature was fixed at 6.4 keV, while the broad line width and normalizations of both the lines were free parameters. With an improvement in goodness of fit ($\chi^2/dof$ = $22.8/17$), the fit with two line components gives narrow and broad line EWs of $28 \pm 20$ eV and $73 \pm 43$ eV, respectively. Using a simple power law continuum parametrization, a broad Gaussian Fe K$\alpha$ line is detected at $\sim$95\% significance ($\Delta\chi^2 \sim$ 9 for two additional parameters), with the width of $\sigma = 0.40^{+0.25}_{-0.14}$ keV. On modeling the broad feature with a {\tt diskline} with fit parameters of a single {\tt diskline} fit, we achieved a similar quality fit ($\chi^2/dof$ = $22.7/17$). The measured EWs of the narrow and diskline components are $39 \pm 14$ eV and $102^{+49}_{-55}$ eV, respectively, while the disk inclination is $44.6^{+3.0}_{-8.4}$ degrees. Figure \ref{fig:cont-pow-nc-disk-rflx}  plots contours (at 68\%, 90\% and 99\% confidence interval) of the {\tt diskline} versus narrow component intensities together with the best fit values derived for this fit ({\tt mo pow+gaussian+diskline}) to the ``averaged ratio flux spectrum''. We also checked the variation in the {\tt diskline} EW by fixing the outer disk radius, $R_{out}$ to 1000 $R_g$ as well to the maximum value of $10^{7} R_g$ allowed by the {\tt diskline} model. The measured {\tt diskline} EW is found to be within the 2 eV difference of the previous value.

Given the robust detection of the narrow Fe K$\alpha$ line in the ``averaged ratio flux spectrum'', a signature of reprocessing in cold, distant  material, we added a neutral reflection component using the {\tt pexrav} in {\tt XSPEC} \citep{Mag1995}. The {\tt pexrav} model computes the spectrum resulting from an exponentially cut-off power law continuum incident on a slab of optically thick, neutral material, accounting for the Compton reflection and bound-free absorption. However, the {\tt pexrav} does not include any fluorescence emission lines. We use the {\tt pexrav} to parametrize the reflected continuum only, as the illuminating primary continuum is separately modeled. In this model, the {\tt pexrav} continuum shape was tied to the mean photon index of the illuminating continuum, $\Gamma$ (fixed at 1.8), abundances were assumed to be Solar, the inclination of the reflector to the observer's line of sight, $cosi$, and the cut-off energy, $foldE$, were fixed at 0.90 and 200 keV, respectively. The solid angle of the reflector $\Omega/2\pi$ was fixed to 1. The reflection fraction of the distant neutral reflector, $R_{Dist}$, is derived from the ratio of the reflection component normalization to the power law normalization. The errors in the $R_{Dist}$ are propagated using the standard error propagation formula \citep{Bev1969}. To account for both the line features we first used two Gaussian lines with parameters set as in the power law continuum modeling. 

Overall, the fit statistic is good ($\chi^2/dof$ = $21.6/17$) with $R_{Dist}$ = $0.27 \pm 0.17$ and the narrow line EW of $28 \pm 20$ eV. With this continuum parametrization, the detection significance of the broad Gaussian line with $\sigma$ = $0.37^{+0.31}_{-0.20}$ is marginal ($\Delta\chi^2 \sim$ 4 for two additional parameters). We therefore inferred an upper limit of 105 eV for the broad line EW. When the broad component is modeled by a {\tt diskline} profile, the fit statistic is similar to the previous fit, with ($\chi^2/dof$ = $21.5/17$) and the EWs of the narrow and diskline are $37 \pm 14$ eV and $<$137 eV, respectively. The fit results are summarized in Table \ref{table:fit-avg-ratio}. The observed variation in the broad line detection significance as a function of the adopted continuum parametrization is noticeable. 

\begin{figure}
\centering
\includegraphics[height=9cm, angle=-90]{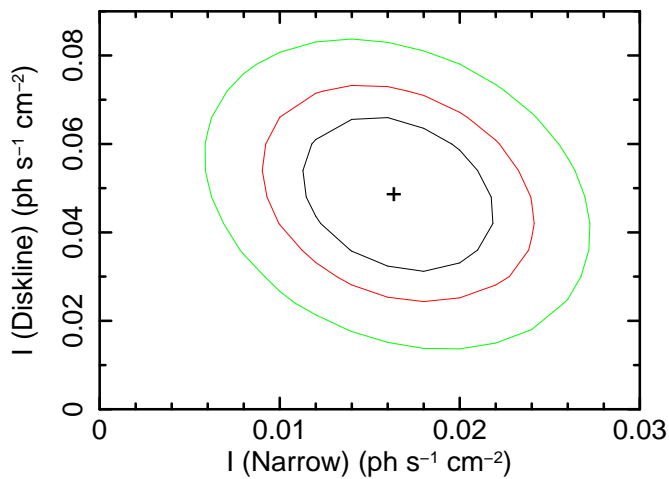}
\caption{Contour plots (68\%, 90\% and 99\%) of the {\tt diskline} versus narrow component intensities for the ``averaged X-ray spectrum''.  A simple power law is used to model the underlying continuum. The cross indicates the best fit values.}
\label{fig:cont-pow-nc-disk-lda}
\end{figure}

\subsection{``Averaged X-ray spectrum'' fitting}\label{sect:Avgspec-fitting}
We have also carried out a detailed spectral analysis of the ``averaged X-ray spectrum'' in the 3--10 keV energy range using the four models discussed above with 2 line components. We mention that in the ``averaged X-ray spectrum'' fitting, the free and fixed parameters of the two lines (parametrized by either Gaussians or a narrow and a {\tt diskline}) are set in accordance with those in the ``averaged ratio flux spectrum'' fitting. First, the underlying continuum was modeled with a power law and two Gaussian lines representing the narrow and broad features were included. The fit statistic was reasonably good ($\chi^2/dof$ = 20.6/17) with the power law photon index, $\Gamma$, of $1.64 \pm 0.03$ and narrow line EW of $28 \pm 10$ eV, while the broad line EW is $92 \pm 35$ eV. The broad Gaussian line is detected at the $\sim$99\% confidence level ($\Delta\chi^2$ = 11 for two additional parameters), being in an excellent agreement to the corresponding fit of the ``averaged ratio flux spectrum''. When the broad feature is modeled by a {\tt diskline}, the fit statistic is excellent ($\chi^2/dof$ = 17.4/17) and we recovered the narrow line EW = $37 \pm 8$ eV, diskline EW = $128 \pm 36$ eV and the disk inclination $i$ = $44.4 \pm 2.4$ degrees. The {\tt diskline} is detected at a robust significance level ($\Delta\chi^2 \sim$ 14 for two additional parameters). Figure \ref{fig:cont-pow-nc-disk-lda} shows contours (at 68\%, 90\% and 99\% confidence interval) of the {\tt diskline} versus narrow component intensities together with the best fit values derived for this fit ({\tt mo pow+gaussian+diskline}) to the ``averaged X-ray spectrum''.  

The additional spectral fits with a simple power law and the reflection component (modeled with {\tt pexrav}), and two line components also give fit parameters fully consistent with those measured in the ``averaged ratio flux spectrum'' fitting, as shown in Table \ref{table:fit-avg-spec}. Also in this case, a strong dependence of the broad line significance on the adopted continuum parametrization is present.  

\subsection{Complex fits}
\label{subsect:complexfits}
We next examine the implications of using a self-consistent reflection spectrum on the broad line detection significance and its measured parameters. We employed the neutral reflection model of \citet{Nandra2007}. This model, known as {\tt pexmon}, enables modeling of the iron lines and Compton reflection continuum in a  self-consistent manner. In addition, the model includes a neutral Fe K$\beta$ line (7.05 keV) with a flux of 11.3 \% of the Fe K$\alpha$ line, a Ni K$\alpha$ with 5 \% of the flux and the Fe K$\alpha$ Compton shoulder. 

\begin{figure}
\centering
\includegraphics[height=6cm]{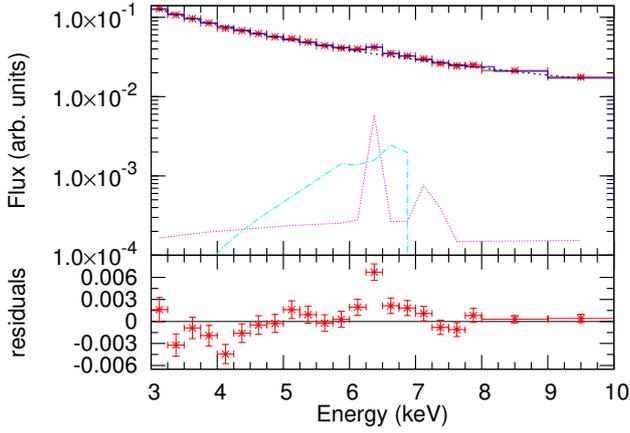}
\caption{{\it Top panel} shows spectral fit to the ``averaged ratio flux spectrum'' using a model including a power law, a distant neutral reflector and a diskline ({\tt mo pow+pexmon+diskline}) along with the model components. Residuals (with the {\tt pexmon} and {\tt diskline} normalizations set to zero for illustration purposes) are shown in the {\it bottom panel}.}
\label{fig:am-ratio-cmplx-fit1}
\end{figure}

Fitting the ``averaged ratio flux spectrum'' with a power law and a distant neutral reflector ({\tt pexmon}) yielded an acceptable fit ($\chi^2/dof$ =  $33.6/20$). However, the fit leaves residuals around the Fe core indicative of the presence of a broad component. When this residual feature is modeled with a {\tt diskline} with parameters shown in Table \ref{table:complexfits}, the fit improves significantly ($\Delta\chi^2 \sim 11.0$). The disk inclination is constrained to be $37.9^{+7.6}_{-2.8}$ degrees, while the {\tt diskline} EW is $89^{+50}_{-38}$ eV. We determined the reflection fraction $R_{Dist} = 0.28 \pm 0.06$ for the reflector. The {\it top panel} of Fig.\ref{fig:am-ratio-cmplx-fit1} shows the spectral fit ({\tt mo pow+pexmon+diskline}) to the ``averaged ratio flux spectrum''. The residuals (with the {\tt pexmon} and {\tt diskline} normalizations set to zero for the sake of illustration) are shown in the {\it bottom panel}. 

The broad residual feature was also parametrized using an emission line profile from an accretion disk around a maximally rotating black hole \citep[{\tt laor},][]{Laor1991}. The peak energy, $E_{Disk}$ was fixed at 6.4 keV, while the inner disk radius, $R_{in}$ and outer disk radius, $R_{out}$, were kept at the default values of 1.24 and 400 $R_g$, respectively, where $R_g$ is the gravitational radius. The emissivity index, $\beta$, was fixed to $2.5$. The disk inclination to the observer, $i$, and the line normalization were allowed to vary. This spectral characterization ({\tt mo pow+pexmon+laor}) yielded an acceptable fit ($\chi^2/dof = 22.1/18$) and we recovered the {\tt laor} line EW = $112^{+50}_{-47}$ eV, the disk inclination $i$ = $43.9^{+3.7}_{-6.1}$ degrees and the reflection fraction $R_{Dist} = 0.26 \pm 0.05$ for the reflector. The fit results are presented in Table \ref{table:complexfits}.
We note that the parameters of the illuminating power law continuum (e.g. photon index $\Gamma$ and normalization) and {\tt pexmon} (e.g. photon index $\Gamma$, cut-off energy, Fe abundance, normalization and the derivation of the reflection fraction $R_{Dist}$ etc.) were treated as in the {\tt pexrav} fits, with the only difference of inclination of the {\tt pexmon} now fixed to 60 degrees.

\begin{figure}
\centering
\includegraphics[height=6cm]{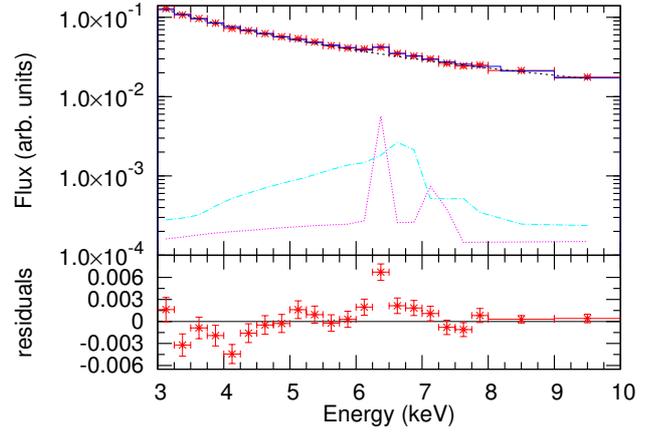} 
\caption{{\it Top panel} shows the spectral fit to the ``averaged ratio flux spectrum'' using a power law and dual reflection spectra ({\tt mo pow+pexmon+kdblur$\ast$pexmon}) along with the model components. Residuals (with the normalizations of both the reflectors set to zero for illustration purposes) are shown in the {\it bottom panel}.}
\label{fig:am-ratio-cmplx-fit2}
\end{figure}

To model the broad feature self-consistently, we assumed that the accretion disk is in a low state of ionization and refitted the ``averaged ratio flux spectrum'' by replacing the {\tt laor} line with a blurred reflection component. We do this by convolving the {\tt pexmon} model with {\tt kdblur}, a convolution model to smooth a spectrum by relativistic effects from an accretion disk around a rotating black hole \citep{Fabian2002}. The inner radius of the blurred reflector (accretion disk) was fixed to $R_{in}$ = 6 $R_{g}$, whereas the outer radius, $R_{out}$, and the emissivity index, $\beta$, were fixed at 100.0 $R_{g}$ and 3.0, respectively. The free parameters were the normalizations of both the reflectors ({\tt pexmon} and {\tt kdblur$\ast$pexmon}) and the inclination, $i$, of the blurred reflector. The average parameters for this model are presented in Table \ref{table:complexfits}. We find that the blurred reflector covers a solid angle (2$\pi  \times R_{Blur}$) of ($0.80 \pm 0.24$)$\pi$ at the X-ray source, while the distant neutral reflector subtends a solid angle of ($2\pi \times R_{Dist}$) of ($0.54 \pm 0.12$)$\pi$ at the X-ray source. The average inclination derived for the blurred reflection component is  $38.8^{+6.2}_{-4.9}$ degrees. In the {\it top panel} of Fig. \ref{fig:am-ratio-cmplx-fit2} we display the spectral fit to the ``averaged ratio flux spectrum'' using a power law and the dual reflection spectra. The residuals (with the normalizations of both the reflectors set to zero for illustration purposes) are shown in the {\it bottom panel}.

Considering that the inner accretion disk is likely to be ionized, we then attempted to model the broad Fe K feature of the ``averaged ratio flux spectrum'' using the blurred ionized reflection component. For doing this, we convolved the ionized reflection model (denoted by {\tt reflionx} in {\tt XSPEC}\footnote{In this model, an optically thick accretion disk is illuminated by a power law, producing fluorescence emission lines and the reflected continuum. The parameters of the {\tt reflionx} model include the incident power law photon index $\Gamma$, Fe abundance and ionization parameter $\xi$, defined as $\xi = 4\pi F_{Tot}/n_H$, where $F_{Tot}$ is the total illuminating flux and $n_H$ is the density of the reflector.}) of \citet{2005MNRAS.358..211R} with {\tt kdblur}. The incident power law photon index of the {\tt reflionx} model was set equal to that of the primary power law component and the Fe abundance was fixed to 1. The ionization parameter and the normalization of the {\tt reflionx} were allowed to vary. The other model components such as the power law, {\tt pexmon} and {\tt kdblur} etc. were set as in the previous fits. 

This model ({\tt mo pow+pexmon+kdblur$\ast$reflionx}) also produced a reasonably good fit with ($\chi^2/dof$ =  $23.5/17$). However, the $\xi$ of the ionized reflector could not be constrained and we inferred an upper limit of 560 erg cm s$^{-1}$ at 68 \%. We outline the fact that the {\tt reflionx} model does not include a formal parameter for the reflection fraction of the accretion disk and provides only the normalization of the  {\tt reflionx} component. Our measured value of the {\tt reflionx} normalization is $(8.2^{+2.5}_{-7.6}) \times 10^{-4}$. We find that the parameters for the power law, distant reflector represented by {\tt pexmon} and {\tt kdblur} components are fully consistent with those obtained in the fit including the blurred neutral reflection component. The fit results are reported in Table \ref{table:complexfits}. 

In the next fits, we replaced the {\tt kdblur} component with the {\tt kerrconv} relativistic convolution model of \citet{2006ApJ...652.1028B} and investigated the variation in the fit parameters for the model comprising dual neutral reflection components. The {\tt kerrconv} model consists of seven parameters: the emissivity indices for the inner and outer disk separated by a break radius, inner and outer radii for the disk emission, the spin parameter of the black hole and the inclination angle of the disk with respect to our line of sight. For the {\tt kerrconv} component, we assumed a single emissivity index of 3. The black hole spin was fixed to 0, while the inclination angle of the disk was kept as a free parameter. We further fixed the inner radius of the disk to be equal to the radius of the marginally stable orbit, $R_{ms}(=6 R_g$ for a non-rotating black hole), and the outer radius to 400 $R_{ms}$. The best-fitting parameters for this model ({\tt mo pow+pexmon+kerrconv$\ast$pexmon}) are given in Table \ref{table:complexfits}. We report that the fit statistic as well as the fit parameters resulted from this model are in excellent agreement with those found in the fit using {\tt kdblur} convolution model. A similar parametrization is obtained when the black hole spin is fixed to 0.998.  

\begin{figure*}
\centering
\includegraphics[height=6cm]{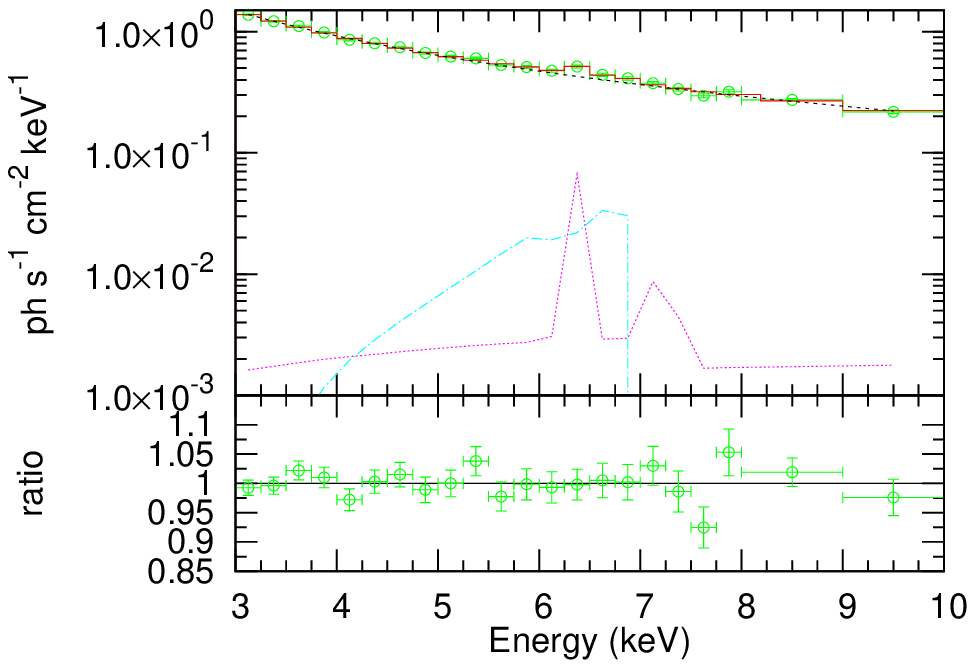} 
\includegraphics[height=6cm]{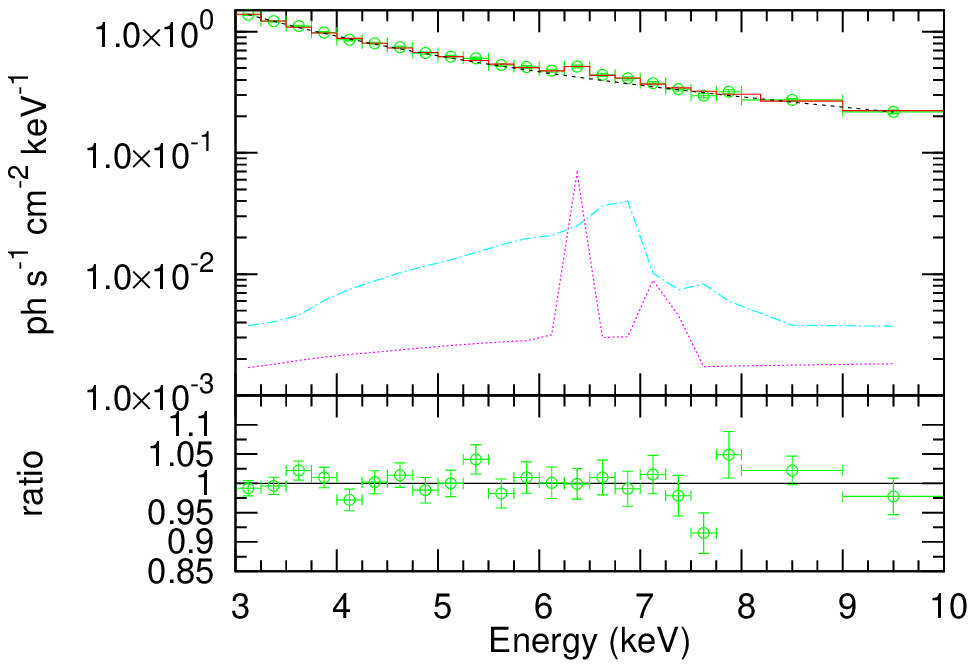} 
\caption{{\it Left panel} shows spectral fits to the ``averaged X-ray spectrum'' using a model including a power law, a distant neutral reflector and a diskline (mo pow+pexmon+diskline). {\it Right panel} displays the spectral fit including a power law and dual reflector spectra (mo pow+pexmon+kdblur$\ast$pexmon). In both figures the {\it bottom panels} show respective data/model ratios.}
\label{fig:am-lda-cmplx-fit1}
\end{figure*}

All the above discussed complex spectral models were also applied to the ``averaged X-ray spectrum''. The fit parameters retrieved from  the ``averaged X-ray spectrum'' spectral fitting are summarized in Table \ref{table:complexfits}. It can be seen that despite the slightly different underlying continuum of the ``averaged X-ray spectrum'', the parameters are fully consistent with those measured from the ``averaged ratio flux spectrum''. For illustration purposes, two spectral fits to the ``averaged X-ray spectrum'' are displayed in Fig.\ref{fig:am-lda-cmplx-fit1}. The {\it left panel} plots spectral fit using a model including a power law, a distant neutral reflector and a diskline ({\tt mo pow+pexmon+diskline}), while the {\it right panel} shows the fit using a power law plus dual reflection spectra ({\tt mo pow+pexmon+kdblur$\ast$pexmon}). 

Finally, considering that the broad line EW or equivalently the reflection fraction of the blurred reflector, $R_{Blur}$, is the critical parameter of our work and many fit parameters (e.g. cut-off energy of the primary power law, Fe abundance, emissivity index, inner and outer radius of the disk etc.) were fixed in our spectral analysis, we investigated the effects of relaxing these fixed parameters on the inferred value of the $R_{Blur}$. We refitted the data by allowing the fixed parameters to vary one by one. We find that the $R_{Blur}$ parameter measured from these new spectral fits is fully consistent with the values reported in Table \ref{table:complexfits} and it is 
$<$1.5 at the 5 sigma level.   

\begin{table*}
\caption{Results of the ``averaged ratio flux spectrum'' spectral fitting in the 3--10 keV}
\begin{tabular}{*{17}{l}}
\hline
\hline
\T Model & \multicolumn{4}{c} {Power law/pexrav} & \multicolumn{3}{c} {Gaussian} &  \multicolumn{7}{c} {Diskline} \\
\cline{2-5} \cline{7-9} \cline{11-16}\\
 & $\Gamma$ & $foldE$ & $R_{Dist}$ & $cos i$ &  &E$_{K\alpha}$ & $\sigma_{K\alpha}$ & EW$_{K\alpha}$ & & $E_{Disk}$ & $R_{in}$ & $R_{out}$ & $\beta$ & $i$  & EW & $\chi^2/dof$ \\
& & (keV) & ($\Omega/2\pi$) & &    &(keV) 		  & (keV) 		& (eV) 		& & (keV) & ($R_{g}$)  & ($R_{g}$)   & 	& (deg) & (eV)& \B \\  
\hline  
\T Pow + Narrow & $1.76\pm0.02$ & $...$ & $...$ & $...$ & &$6.40^\ast$ & $0.01^\ast$ & $45\pm13$ & & $...$ & $...$ & $...$ & $...$ & $...$ & $...$ & $31.3/19$ \B \\

Pow + Broad & $1.77\pm0.02$ & $...$ & $...$ & $...$ & &$6.40^\ast$ & $0.15^{+0.06}_{-0.04}$ & $75\pm40$& & $...$  & $...$ & $...$ & $...$ & $...$ & $...$ & $25.6/18$ \B\\ 

Pow + Disk & $1.78^{n}$ & $...$ & $...$ & $...$ & &$...$ & $...$ & $...$& & $6.40^\ast$  & $6.0^\ast$ & $100.0^\ast$ & $-2.5^\ast$ & $35.3^n$ & $128\pm41$ &  $41.6/18$ \B \\ \hline

\T Pow + Narrow + Broad & $1.78\pm0.02$ & $...$ & $...$ & $...$  & &$6.40^\ast$ & $0.01^\ast$ & $28\pm20$ & & $...$ & $...$ & $...$ & $...$ & $...$ & $...$ & \B \\
& $...$ & $...$ & $...$ & $...$ & &$6.40^\ast$ & $0.40^{+0.25}_{-0.14}$ & $73\pm43$& & $...$  & $...$ & $...$ & $...$ & $...$ & $...$ &  $22.8/17$  \B \\  

Pow + Narrow + Disk & $1.78\pm0.02$ & $...$ & $...$ & $...$ & & $6.40^\ast$ & $0.01^\ast$ & $39\pm14$ & & $6.40^\ast$ & $6.0^\ast$  & $100.0^\ast$ & $-2.5^\ast$ & $44.6^{+3.0}_{-8.4}$ & $102^{+49}_{-55}$ & $22.7/17$ \B \\

Pow + pexrav + Narrow + Broad & $1.8^\ast$ & $200^\ast$ & $0.27\pm0.17$ & $0.9^\ast$ & & $6.40^\ast$ & $0.01^\ast$ & $28\pm20$ & & $...$ & $...$ & $...$ & $...$ & $...$ & $...$ & \B \\
& $...$ & $...$ & $...$ & $...$ & &$6.40^\ast$ & $0.37^{+0.31}_{-0.20}$ & $<105$& & $...$  & $...$ & $...$ & $...$ & $...$ & $...$ &  $21.6/17$ \B \\ 

Pow + pexrav + Narrow + Disk & $1.8^\ast$ & $200^\ast$ & $0.24\pm0.17$ & $0.9^\ast$ & & $6.40^\ast$ & $0.01^\ast$ & $37\pm14$ & & $6.40^\ast$ & $6.0^\ast$  & $100.0^\ast$ & $-2.5^\ast$ & $45.5\pm3.5$ & $<137$ & $21.5/17$ \B \\
\hline  
\end{tabular}
\label{table:fit-avg-ratio}
\\
\\
Notes: All errors and upper limits refer to the 68\% confidence range for a single parameter.\\
In all the spectral fits, the errors in the Fe lines EW are calculated using {\tt XSPEC} {\it eqw with err option}. \\
$^\ast$ denotes fixed parameter.\\
$^n$ Error calculation is not possible due to poor fit. \\
\end{table*}

\begin{table*}
\caption{Results of the ``averaged X-ray spectrum'' spectral fitting in the 3--10 keV}
 \begin{tabular}{*{17}{l}}
\hline
\hline
\T Model & \multicolumn{4}{c} {Power law/pexrav} & \multicolumn{3}{c} {Gaussian} &  \multicolumn{7}{c} {Diskline} \\
\cline{2-5} \cline{7-9} \cline{11-16}\\
 & $\Gamma$ & $foldE$ & $R_{Dist}$ & $cos i$ &  &E$_{K\alpha}$ & $\sigma_{K\alpha}$ & EW$_{K\alpha}$ & & $E_{Disk}$ & $R_{in}$ & $R_{out}$ & $\beta$ & $i$  & EW & $\chi^2/dof$ \\
& & (keV) & ($\Omega/2\pi$) & &    &(keV) 		  & (keV) 		& (eV) 		& & (keV) & ($R_{g}$)  & ($R_{g}$)   & 	& (deg) & (eV) & \B \\  
\hline
\T Pow + Narrow + Broad & $1.64\pm0.02$ & $...$ & $...$ & $...$& & $6.40^\ast$ & $0.01^\ast$ & $28\pm10$ & & $...$ & $...$ & $...$ & $...$ & $...$ & $...$ & \B  \\
& $...$ & $...$ & $...$ & $...$ & & $6.40^\ast$ & $0.49^{+0.22}_{-0.14}$ & $92\pm35$ & & $...$ & $...$ & $...$ & $...$ & $...$ & $...$ & $20.6/17$\B  \\

Pow + Narrow + Disk & $1.64\pm0.02$ & $...$ & $...$ & $...$ & & $6.40^\ast$ & $0.01^\ast$ & $37\pm8$ & & $6.40^\ast$ & $6.0^\ast$  & $100.0^\ast$ & $-2.5^\ast$ & $44.4\pm2.4$ & $128\pm36$ & $17.4/17$ \B \\

Pow + pexrav + Narrow + Broad & $1.67^\ast$ & $200^\ast$ & $0.29\pm0.18$ & $0.9^\ast$ & & $6.40^\ast$ & $0.01^\ast$ & $29\pm10$ & & $...$ & $...$ & $...$ & $...$ & $...$ & $...$ & \B  \\
  & $...$ & $...$ & $...$ & $...$ & & $6.40^\ast$ & $0.52^{+0.29}_{-0.18}$ & $78\pm40$ & & $...$ & $...$ & $...$ & $...$ & $...$ & $...$ & $20.2/17$\B  \\

Pow + pexrav + Narrow + Disk & $1.67^\ast$ & $200^\ast$ & $0.25\pm0.17$ & $0.9^\ast$ & & $6.40^\ast$ & $0.01^\ast$ & $36\pm8$ & & $6.40^\ast$ & $6.0^\ast$  & $100.0^\ast$ & $-2.5^\ast$ & $45.1^{+3.8}_{-2.3}$ & $114\pm41$ & $17.4/17$ \B \\

\hline  
\end{tabular}
\label{table:fit-avg-spec}
\\
\\
All errors and upper limits are quoted at the 68\% confidence range for a single parameter.\\
In all the spectral fits, the errors in the Fe lines EW are calculated using {\tt XSPEC} {\it eqw with err option}. \\
$^\ast$ denotes fixed parameter.\\
\end{table*}

\begin{table*}
\caption{Results of the ``averaged ratio flux spectrum'' and ``averaged X-ray spectrum'' spectral fitting in the 3--10 keV using complex models}
\begin{tabular}{*{14}{l}}
\hline
\hline
\T Model & Power law & \multicolumn{3}{c} {Distant Reflector} & \multicolumn{8}{c} {Diskline/Blurred Reflector} \\
\cline{3-5} \cline{7-13}\\
 & $\Gamma$ & $foldE$ & $R_{Dist}$ & $i$ & & $E_{Disk}$ & $R_{in}$ & $R_{out}$ & $\beta$ & $i$  & EW & $R_{Blur}$ & $\chi^2/dof$ \\
 &          & (keV)     & ($\Omega/2\pi$) & (deg) &  &  (keV)        & ($R_{g}$) & ($R_{g}$)   &         & (deg)  & (eV)   & ($\Omega/2\pi$) & \B \\ \hline
\multicolumn{13}{c} {``averaged ratio flux spectrum''} \T \B \\ \hline
\T Pow + pexmon + Disk & $1.8^\ast$ & $200^\ast$ & $0.28\pm0.06$ & $60.0^\ast$ & & $6.40^\ast$ & $6.0^\ast$  & $100.0^\ast$ & $-2.5^\ast$ & $37.9^{+7.6}_{-2.8}$ & $89^{+50}_{-38}$ & $...$ & $22.7/18$ \B  \\

Pow + pexmon + Laor & $1.8^\ast$ & $200^\ast$ & $0.26\pm0.05$ & $60.0^\ast$ & & $6.40^\ast$ & $1.24^\ast$  & $400.0^\ast$ & $2.5^\ast$ & $43.9^{+3.7}_{-6.1}$ & $112^{+50}_{-47}$ & $...$ & $22.1/18$ \B \\

Pow + pexmon + kdblur$\ast$pexmon & $1.8^\ast$ & $200^\ast$ & $0.27\pm0.06$ & $60.0^\ast$ & & $...$ & $6.0^\ast$  & $100.0^\ast$ & $3.0^\ast$ & $38.8^{+6.2}_{-4.9}$ & $...$ & $0.40\pm0.12$ & $22.9/18$ \B \\  

Pow + pexmon + kdblur$\ast$reflionx & $1.8^\ast$ & $200^\ast$ & $0.27\pm0.06$ & $60.0^\ast$ & & $...$ & $6.0^\ast$ & $400.0^\ast$ & $3.0^\ast$ & $39.7^{+5.9}_{-7.1}$ & $...$ & $...$ & $23.5/17$ \B \\ 

Pow + pexmon + kerrconv$\ast$pexmon & $1.8^\ast$ & $200^\ast$ & $0.27\pm0.06$ & $60.0^\ast$ & & $...$ & $ 6.0^\ast$ & $2400.0^\ast$ & $3.0^\ast$ & $43.9^{+2.8}_{-8.4}$ & $...$ & $0.40\pm0.12$ & $23.1/18$ \B \\ \hline

\multicolumn{13}{c} {``averaged X-ray spectrum''} \T \B \\ \hline
\T Pow + pexmon + Disk & $1.65\pm0.02$ & $200^\ast$ & $0.24\pm0.06$ & $60.0^\ast$ & & $6.40^\ast$ & $6.0^\ast$  & $100.0^\ast$ & $-2.5^\ast$ & $38.5^{+6.8}_{-2.4}$ & $102^{+35}_{-28}$ & $...$ & $17.5/17$ \B \\

Pow + pexmon + Laor & $1.65\pm0.02$ & $200^\ast$ & $0.23\pm0.06$ & $60.0^\ast$ & & $6.40^\ast$ & $1.24^\ast$  & $400.0^\ast$ & $2.7\pm0.6$ & $43.9^{+3.3}_{-4.0}$ & $167^{+48}_{-61}$ & $...$ & $16.6/16$ \B \\

Pow + pexmon + kdblur$\ast$pexmon & $1.67\pm0.02$ & $200^\ast$ & $0.25\pm0.06$ & $60.0^\ast$ & & $...$ & $6.0^\ast$  & $100.0^\ast$ & $3.0^\ast$ & $40.8\pm4.1$ & $...$ & $0.49\pm0.15$ & $18.6/17$ \B \\

Pow + pexmon + kdblur$\ast$reflionx & $1.67\pm0.06$ & $200^\ast$ & $0.27\pm0.07$ & $60.0^\ast$ & & $...$ & $1.24^\ast$  & $400.0^\ast$ & $3.0^\ast$ & $33.8^{+4.9}_{-5.3}$ & $...$ & $...$ & $18.8/16$ \B \\

Pow + pexmon + kerrconv$\ast$pexmon & $1.66\pm0.02$ & $200^\ast$ & $0.24\pm0.05$ & $60.0^\ast$ & & $...$ & $6.0^\ast$  & $2400.0^\ast$ & $3.0^\ast$ & $43.8^{+2.6}_{-6.0}$ & $...$ & $0.50\pm0.16$ & $18.3/17$ \B \\ \hline
\hline  
\end{tabular}
\label{table:complexfits}
\\
\\
Notes: All errors and upper limits refer to the 68\% confidence range for a single parameter.\\
In all the spectral fits, the errors in the Fe lines EW are calculated using {\tt XSPEC} {\it eqw with err option}. \\
In the fits including {\tt kerrconv} component, the black hole spin is fixed to 0.\\ 
$^\ast$ denotes fixed parameter.\\
\end{table*}

\section{Discussion and perspectives}\label{sect:Discussion}

It is critical to have multiple independent methods to measure the average Fe K$\alpha$ line properties. We have determined the Fe line parameters using two methods described in Sect. \ref{sect:RF-stacking-proc}. We find that the average narrow core EW for the whole sample comprising 248 AGNs is $\sim$30 eV and is stable in various continuum models employed for the ``averaged ratio flux spectrum''  and ``averaged X-ray spectrum'' (Tables \ref{table:fit-avg-ratio} and \ref{table:fit-avg-spec}). We have demonstrated that the broad line parameters and its detection significance are highly sensitive to the adopted continuum modeling. 

\begin{figure}
\centering
\includegraphics[height=6cm]{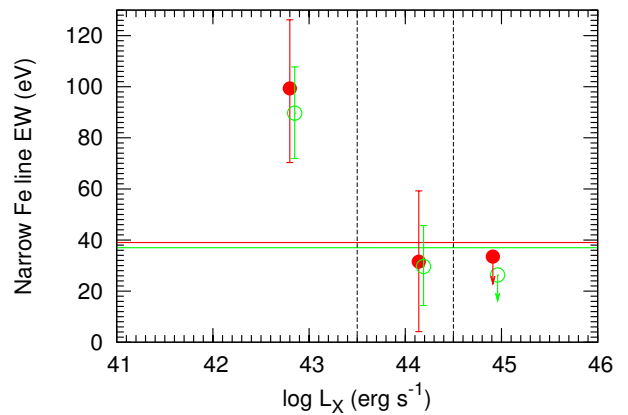}
\caption{Equivalent width of the narrow Fe K$\alpha$ line measured from the spectral fitting (mo pow+gaussian+diskline) of the ``averaged ratio flux spectrum'' (red) and ``averaged X-ray spectrum'' (green) of the total sample comprising 248 AGNs (horizontal lines) and the sub-samples used in 3 hard X-ray luminosity bins illustrated by the dotted vertical lines. The EW is plotted against the median luminosity in  each bin. Errors and upper limits are drawn at the 68$\%$ confidence level.}
 \label{fig:comp-nc-EW-LBINS}
 \end{figure}

In order to examine the average Fe line properties as a function of the hard X-ray luminosity (2--10 keV) and also to investigate the variation in the line parameters obtained from the two methods we splitted the sample in 3 luminosity intervals (L1, $41 \leq logL_X \leq 43.5$ erg s$^{-1}$; L2, $43.5 < logL_X \leq 44.5$ erg s$^{-1}$; L3, $44.5 < logL_X \leq 46.0$ erg s$^{-1}$). We then computed the ''averaged ratio flux spectra'' from the averaged ratios and ``averaged X-spectra'' in these luminosity bins and performed their spectral analysis. Fig. \ref{fig:comp-nc-EW-LBINS} plots the narrow Fe K$\alpha$ EWs measured from the spectral fitting of the ``averaged ratio flux spectrum'' (red) and ``averaged X-ray spectrum'' (green) against the X-ray luminosity in the 2--10 keV band for the whole sample of 248 AGNs (horizontal lines) and its sub-samples used for luminosity bins. An excellent agreement between the narrow line EWs determined from the two independent methods is evident. The narrow line EW decreases as the X-ray luminosity increases, a trend designated as the ``X-ray Baldwin effect'' or ``Iwasawa-Taniguchi effect'', first reported by \citet{IT1993}. We do not concentrate on quantifying the X-ray luminosity dependence of the narrow line EW as it has been discussed in our previous work (see CH10). However, we point out that the narrow line EWs derived in the rest-frame are consistent with the ones measured in the observed frame. The neutral, narrow Fe K$\alpha$ fluorescent emission line, peaking at 6.4 keV is reported to be a common and dominant feature in the X-ray spectra of local and distant AGNs \citep{Yaqoob2004, Nandra2007, Shu2010, Corral2008, Chaudhary2010}. Our analysis further confirms this important finding.  

\begin{figure}
\centering
\includegraphics[height=6cm]{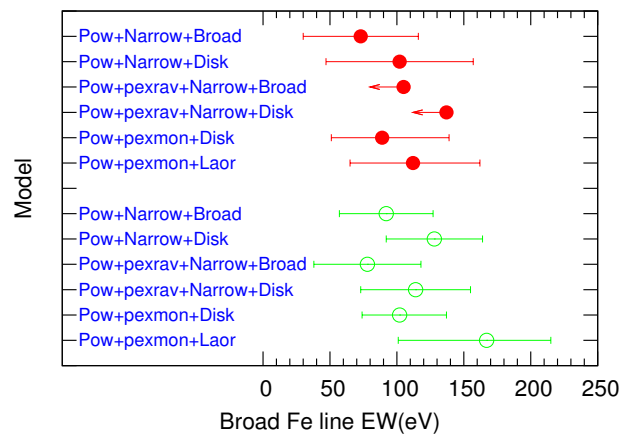} 
\caption{Comparison of the equivalent width of the broad Fe K$\alpha$ line measured from the spectral fitting of the ``averaged ratio flux spectrum (red) and ``averaged X-ray spectrum'' (green) of the total sample comprising 248 AGNs. Respective deterministic model is listed in blue. Errors and upper limits are drawn at the 68$\%$ confidence level.}
\label{fig:comp-br-EW-avg-RatiovsSpec}
\end{figure}

When the broad line parameters such as its shape, EW and its detection significance in the ``averaged ratio flux spectrum'' and ``averaged X-ray spectrum'' of the total sample of 248 AGNs are considered, we notice that these parameters are dependent on the assumed continuum and adopted stacking method as illustrated in Fig. \ref{fig:comp-br-EW-avg-RatiovsSpec}. In particular, despite having a well defined sample with reasonable statistics (net counts $\sim$ 198 000) in the ``averaged X-ray spectrum'', we do not detect a clear extended red-wing, and the measured EW of the broad feature is always lower than 170 eV, in a good agreement with the recent studies of the average Fe K emission from AGNs \citep{Gu2006, Corral2008, Longinotti2008}. Although we do not present the details of the fits in the 2--10 keV band, we have performed those fits. We note that the broad line EW measured from the fits performed in the whole 2--10 keV range is always lower than 160 eV, which is statistically consistent with the value found in the 3--10 keV band.

Our derived broad line EW is significantly lower than the $\sim$500 eV reported in the stacked spectrum of AGNs in the Lockman Hole \citep[][ hereafter S05]{Alina2005}. We briefly discuss possible causes for the difference. Considering that the fluorescent line profile originating from an accretion disk is a strong function of the disk ionization state \citep{Fabian2000} and hence of the X-ray luminosity, we first investigated if the observed discrepancy in the Fe K line EW is caused by the different luminosity distribution of the two samples. The luminosities of the Lockman Hole sample were taken from \citet{Main2002}. The 2--10 keV X-ray luminosity distributions of 53 Type 1 and 41 Type 2 AGNs used in S05 are very similar, with the mean luminosity of 10$^{44.0}$ ergs s$^{-1}$. The mean 2--10 keV X-ray luminosity of our sample is 10$^{44.2}$ ergs s$^{-1}$. We conclude that the luminosity difference does not contribute to the difference of the line strengths. Our measured broad line EW determined from a simple power law continuum parametrization is much lower than the associated value derived in S05. This implies that the continuum modeling is also not the key parameter behind the discrepant results. The observed discrepancy is most likely due to the difference in the adopted averaging techniques.    

We also performed complex fits presented in Sect. \ref{subsect:complexfits} by modeling the ``averaged ratio flux spectrum'' and ``averaged X-ray spectrum'' with dual reflection models (a distant reflector and a disk reflector) handling emission lines and the Compton reflection self-consistently. Two types of reflection models (i.e. neutral and ionized) were applied to the data. We stress that from the statistical view point, in both the explored averaging procedures, all models including two line components or dual reflection spectra provide good description of the data. Interestingly, characterization of the distant reflector with the {\tt pexmon} model results in a high detection significance of the observed broad feature as compared to the {\tt pexrav}. Given the available data quality and the well known degeneracy of accretion disk parameters, we could not derive all fit parameters independently. However, we emphasize that we have explored the parameter space in detail by relaxing the fixed parameters one by one. We note that the measured parameters from these fits are fully consistent with the values listed in Table \ref{table:complexfits}. In particular, based on this parameter space exploration, we can safely say that the $R_{Blur}$ parameter determined from our data is $<$1.5 at the 5 sigma level. Furthermore, the derived average parameters of the dual reflectors such as the solid angle covered by the reflectors at the X-ray source and the inclination etc. agree well with the average values reported in the local AGNs samples with high counting statistics \citep{Nandra2007, Dela2010}.     

According to the detailed calculations of \citet{George1991}, the predicted EW of the Fe K$\alpha$ arising from a centrally illuminated cold disk is 130--140 eV for a continuum with $\Gamma$ in the range 1.8--1.9 and an inclination of 40 degrees. Our observed mean {\tt diskline} EW of $\sim$100 eV agrees well with the theoretical prediction. Our results are also consistent with predictions of the relativistic Fe K$\alpha$ line intensities from the integrated spectra of AGNs presented by \citep[][see Figure 1]{Ballan2010}. In particular, the measured broad line EW and accretion disk reflection fraction ($R_{Blur}$ $\sim$ 0.5) in our data are consistent with \citet{Ballan2010} predictions, despite the slightly different assumptions on the relativistic profile for the line ({\tt laor2} versus {\tt diskline}). 

In future work we plan to substantially improve the statistics by assembling larger AGNs samples. In particular, including MOS1 and MOS2 data and collecting multiwavelength data to study the broad Fe line properties as a function of the source physical properties, such as the hard X-ray luminosity, black hole mass and accretion rate \citep[see e.g. ][]{Ballan2010}. Further progress in exploiting the full diagnostic potential of the Fe K$\alpha$ line in measuring black hole properties (e.g. mass and spin) requires more data with newer, more sensitive instruments with very high effective area and broad band coverage. The next generation X-ray observatory, ATHENA will provide these capabilities. Owing to its huge effective area, ATHENA will enable the measurement and characterization of broad lines in individual AGNs covering a broad redshift range ($z$ = 1--2). 

\acknowledgement
Our results are based on observations obtained with XMM-{\it Newton}, an ESA science mission with instruments and contributions directly funded by ESA member states and the USA (NASA). The XMM-{\it Newton} project is supported by the Bundesministerium f\"ur Wirtschaft und Technologie/Deutsches Zentrum f\"ur Luft und Raumfahrt (BMWI/DLR, FKZ 50 OX 0001), the Max-Planck Society and the Heidenhain-Stiftung and also by PPARC, CEA, CNES, and ASI. This research has made use of the NASA/IPAC Extragalactic Database (NED) which is operated by the Jet Propulsion Laboratory, California Institute of Technology, under contract with the National Aeronautics and Space Administration. PC acknowledges support from and participation in the International Max-Planck Research School on Astrophysics at the Ludwig-Maximilians University. GH acknowledges support from the German Deutsche Forschungsgemeinschaft, DFG, Leibniz Prize (FKZ HA 1850/28–1). AC acknowledges support from the Italian Space Agency (ASI) under the contracts ASI-INAF I/088/06/0 and I/009/10/0. The authors would like to thank the anonymous referee for the encouraging report and useful comments, which improved the paper significantly. We thank Kazushi Iwasawa for providing many details of his stacking method, Andy Fabian for an enlightening discussion.  

\bibliographystyle{aa}
\bibliography{references}

\begin{thebibliography}{41}
\expandafter\ifx\csname natexlab\endcsname\relax\def\natexlab#1{#1}\fi

\bibitem[{{Antonucci}(1993)}]{Antonucci1993}
{Antonucci}, R. 1993, \araa, 31, 473

\bibitem[{{Arnaud}(1996)}]{Arnaud1996}
{Arnaud}, K.~A. 1996, in Astronomical Society of the Pacific Conference Series,
  Vol. 101, Astronomical Data Analysis Software and Systems V, ed. G.~H.
  {Jacoby} \& J.~{Barnes}

\bibitem[{{Ballantyne}(2010)}]{Ballan2010}
{Ballantyne}, D.~R. 2010, \apjl, 716, L27

\bibitem[{{Bevington}(1969)}]{Bev1969}
{Bevington}, P.~R. 1969, {Data reduction and error analysis for the physical
  sciences}, ed. {Bevington, P.~R.}

\bibitem[{{Bianchi} {et~al.}(2007){Bianchi}, {Guainazzi}, {Matt}, \& {Fonseca
  Bonilla}}]{Bianchi2007}
{Bianchi}, S., {Guainazzi}, M., {Matt}, G., \& {Fonseca Bonilla}, N. 2007,
  \aap, 467, L19

\bibitem[{{Brenneman} \& {Reynolds}(2006)}]{2006ApJ...652.1028B}
{Brenneman}, L.~W. \& {Reynolds}, C.~S. 2006, \apj, 652, 1028

\bibitem[{{Brusa} {et~al.}(2005){Brusa}, {Gilli}, \& {Comastri}}]{Brusa2005}
{Brusa}, M., {Gilli}, R., \& {Comastri}, A. 2005, \apjl, 621, L5

\bibitem[{{Cash}(1979)}]{Cash1979}
{Cash}, W. 1979, \apj, 228, 939

\bibitem[{{Chaudhary} {et~al.}(2010){Chaudhary}, {Brusa}, {Hasinger},
  {Merloni}, \& {Comastri}}]{Chaudhary2010}
{Chaudhary}, P., {Brusa}, M., {Hasinger}, G., {Merloni}, A., \& {Comastri}, A.
  2010, \aap, 518, A58

\bibitem[{{Corral} {et~al.}(2008){Corral}, {Page}, {Carrera}, {Barcons},
  {Mateos}, {Ebrero}, {Krumpe}, {Schwope}, {Tedds}, \& {Watson}}]{Corral2008}
{Corral}, A., {Page}, M.~J., {Carrera}, F.~J., {et~al.} 2008, \aap, 492, 71

\bibitem[{{de La Calle P{\'e}rez} {et~al.}(2010){de La Calle P{\'e}rez},
  {Longinotti}, {Guainazzi}, {Bianchi}, {Dov{\v c}iak}, {Cappi}, {Matt},
  {Miniutti}, {Petrucci}, {Piconcelli}, {Ponti}, {Porquet}, \&
  {Santos-Lle{\'o}}}]{Dela2010}
{de La Calle P{\'e}rez}, I., {Longinotti}, A.~L., {Guainazzi}, M., {et~al.}
  2010, \aap, 524, A50

\bibitem[{{Fabian} {et~al.}(2000){Fabian}, {Iwasawa}, {Reynolds}, \&
  {Young}}]{Fabian2000}
{Fabian}, A.~C., {Iwasawa}, K., {Reynolds}, C.~S., \& {Young}, A.~J. 2000,
  \pasp, 112, 1145

\bibitem[{{Fabian} \& {Miniutti}(2005)}]{Fabian2005}
{Fabian}, A.~C. \& {Miniutti}, G. 2005, ArXiv Astrophysics e-prints

\bibitem[{{Fabian} {et~al.}(1989){Fabian}, {Rees}, {Stella}, \&
  {White}}]{Fabian1989}
{Fabian}, A.~C., {Rees}, M.~J., {Stella}, L., \& {White}, N.~E. 1989, \mnras,
  238, 729

\bibitem[{{Fabian} {et~al.}(2002){Fabian}, {Vaughan}, {Nandra}, {Iwasawa},
  {Ballantyne}, {Lee}, {De Rosa}, {Turner}, \& {Young}}]{Fabian2002}
{Fabian}, A.~C., {Vaughan}, S., {Nandra}, K., {et~al.} 2002, \mnras, 335, L1

\bibitem[{{George} \& {Fabian}(1991)}]{George1991}
{George}, I.~M. \& {Fabian}, A.~C. 1991, \mnras, 249, 352

\bibitem[{{Guainazzi} {et~al.}(2006){Guainazzi}, {Bianchi}, \& {Dov{\v
  c}iak}}]{Gu2006}
{Guainazzi}, M., {Bianchi}, S., \& {Dov{\v c}iak}, M. 2006, Astronomische
  Nachrichten, 327, 1032

\bibitem[{{Guilbert} \& {Rees}(1988)}]{Guilbert1988}
{Guilbert}, P.~W. \& {Rees}, M.~J. 1988, \mnras, 233, 475

\bibitem[{{Iwasawa} \& {Taniguchi}(1993)}]{IT1993}
{Iwasawa}, K. \& {Taniguchi}, Y. 1993, \apjl, 413, L15

\bibitem[{{Laor}(1991)}]{Laor1991}
{Laor}, A. 1991, \apj, 376, 90

\bibitem[{{Lightman} \& {White}(1988)}]{Lightman1988}
{Lightman}, A.~P. \& {White}, T.~R. 1988, \apj, 335, 57

\bibitem[{{Longinotti} {et~al.}(2008){Longinotti}, {de La Calle}, {Bianchi},
  {Guainazzi}, \& {Dov{\v c}iak}}]{Longinotti2008}
{Longinotti}, A.~L., {de La Calle}, I., {Bianchi}, S., {Guainazzi}, M., \&
  {Dov{\v c}iak}, M. 2008, \memsai, 79, 259

\bibitem[{{Magdziarz} \& {Zdziarski}(1995)}]{Mag1995}
{Magdziarz}, P. \& {Zdziarski}, A.~A. 1995, \mnras, 273, 837

\bibitem[{{Mainieri} {et~al.}(2002){Mainieri}, {Bergeron}, {Hasinger},
  {Lehmann}, {Rosati}, {Schmidt}, {Szokoly}, \& {Della Ceca}}]{Main2002}
{Mainieri}, V., {Bergeron}, J., {Hasinger}, G., {et~al.} 2002, \aap, 393, 425

\bibitem[{{Mao} {et~al.}(2010){Mao}, {Hu}, {Wang}, {Zhao}, \&
  {Zhang}}]{Mao2010}
{Mao}, W., {Hu}, C., {Wang}, J., {Zhao}, G., \& {Zhang}, S. 2010, Research in
  Astronomy and Astrophysics, 10, 905

\bibitem[{{Matt} {et~al.}(1991){Matt}, {Perola}, \& {Piro}}]{Matt1991}
{Matt}, G., {Perola}, G.~C., \& {Piro}, L. 1991, \aap, 247, 25

\bibitem[{{Miller}(2007)}]{Miller2007}
{Miller}, J.~M. 2007, \araa, 45, 441

\bibitem[{{Miller} {et~al.}(2009){Miller}, {Turner}, \& {Reeves}}]{Miller2009}
{Miller}, L., {Turner}, T.~J., \& {Reeves}, J.~N. 2009, \mnras, 399, L69

\bibitem[{{Nandra} {et~al.}(1997){Nandra}, {George}, {Mushotzky}, {Turner}, \&
  {Yaqoob}}]{Nandra1997}
{Nandra}, K., {George}, I.~M., {Mushotzky}, R.~F., {Turner}, T.~J., \&
  {Yaqoob}, T. 1997, \apjl, 488, L91

\bibitem[{{Nandra} {et~al.}(2007){Nandra}, {O'Neill}, {George}, \&
  {Reeves}}]{Nandra2007}
{Nandra}, K., {O'Neill}, P.~M., {George}, I.~M., \& {Reeves}, J.~N. 2007,
  \mnras, 382, 194

\bibitem[{{Page} {et~al.}(2004){Page}, {O'Brien}, {Reeves}, \&
  {Turner}}]{Page2004}
{Page}, K.~L., {O'Brien}, P.~T., {Reeves}, J.~N., \& {Turner}, M.~J.~L. 2004,
  \mnras, 347, 316

\bibitem[{{Patrick} {et~al.}(2010){Patrick}, {Reeves}, {Porquet}, {Markowitz},
  {Lobban}, \& {Terashima}}]{Patrick2010}
{Patrick}, A.~R., {Reeves}, J.~N., {Porquet}, D., {et~al.} 2010, ArXiv e-prints

\bibitem[{{Reeves} {et~al.}(2006){Reeves}, {Fabian}, {Kataoka}, {Kunieda},
  {Markowitz}, {Miniutti}, {Okajima}, {Serlemitsos}, {Takahashi}, {Terashima},
  \& {Yaqoob}}]{Reeves2006}
{Reeves}, J.~N., {Fabian}, A.~C., {Kataoka}, J., {et~al.} 2006, Astronomische
  Nachrichten, 327, 1079

\bibitem[{{Reynolds} \& {Nowak}(2003)}]{Reynolds2003}
{Reynolds}, C.~S. \& {Nowak}, M.~A. 2003, \physrep, 377, 389

\bibitem[{{Ross} \& {Fabian}(2005)}]{2005MNRAS.358..211R}
{Ross}, R.~R. \& {Fabian}, A.~C. 2005, \mnras, 358, 211

\bibitem[{{Shu} {et~al.}(2010){Shu}, {Yaqoob}, \& {Wang}}]{Shu2010}
{Shu}, X.~W., {Yaqoob}, T., \& {Wang}, J.~X. 2010, \apjs, 187, 581

\bibitem[{{Streblyanska} {et~al.}(2005){Streblyanska}, {Hasinger},
  {Finoguenov}, {Barcons}, {Mateos}, \& {Fabian}}]{Alina2005}
{Streblyanska}, A., {Hasinger}, G., {Finoguenov}, A., {et~al.} 2005, \aap, 432,
  395

\bibitem[{{Tanaka} {et~al.}(1995){Tanaka}, {Nandra}, {Fabian}, {Inoue},
  {Otani}, {Dotani}, {Hayashida}, {Iwasawa}, {Kii}, {Kunieda}, {Makino}, \&
  {Matsuoka}}]{Tanaka1995}
{Tanaka}, Y., {Nandra}, K., {Fabian}, A.~C., {et~al.} 1995, \nat, 375, 659

\bibitem[{{Turner} \& {Miller}(2009)}]{Turner2009}
{Turner}, T.~J. \& {Miller}, L. 2009, \aapr, 17, 47

\bibitem[{{Watson} {et~al.}(2009){Watson}, {Schr{\"o}der}, {Fyfe}, {Page},
  {Lamer}, {Mateos}, {Pye}, {Sakano}, {Rosen}, {Ballet}, {Barcons}, {Barret},
  {Boller}, {Brunner}, {Brusa}, {Caccianiga}, {Carrera}, {Ceballos}, {Della
  Ceca}, {Denby}, {Denkinson}, {Dupuy}, {Farrell}, {Fraschetti}, {Freyberg},
  {Guillout}, {Hambaryan}, {Maccacaro}, {Mathiesen}, {McMahon}, {Michel},
  {Motch}, {Osborne}, {Page}, {Pakull}, {Pietsch}, {Saxton}, {Schwope},
  {Severgnini}, {Simpson}, {Sironi}, {Stewart}, {Stewart}, {Stobbart}, {Tedds},
  {Warwick}, {Webb}, {West}, {Worrall}, \& {Yuan}}]{Watson2009}
{Watson}, M.~G., {Schr{\"o}der}, A.~C., {Fyfe}, D., {et~al.} 2009, \aap, 493,
  339

\bibitem[{{Yaqoob} \& {Padmanabhan}(2004)}]{Yaqoob2004}
{Yaqoob}, T. \& {Padmanabhan}, U. 2004, \apj, 604, 63

\end{thebibliography}

\begin{appendix}
\section{Spectral fits}

\begin{figure*}[h]
\centering
\begin{tabular}{cc}
\includegraphics[height=5cm]{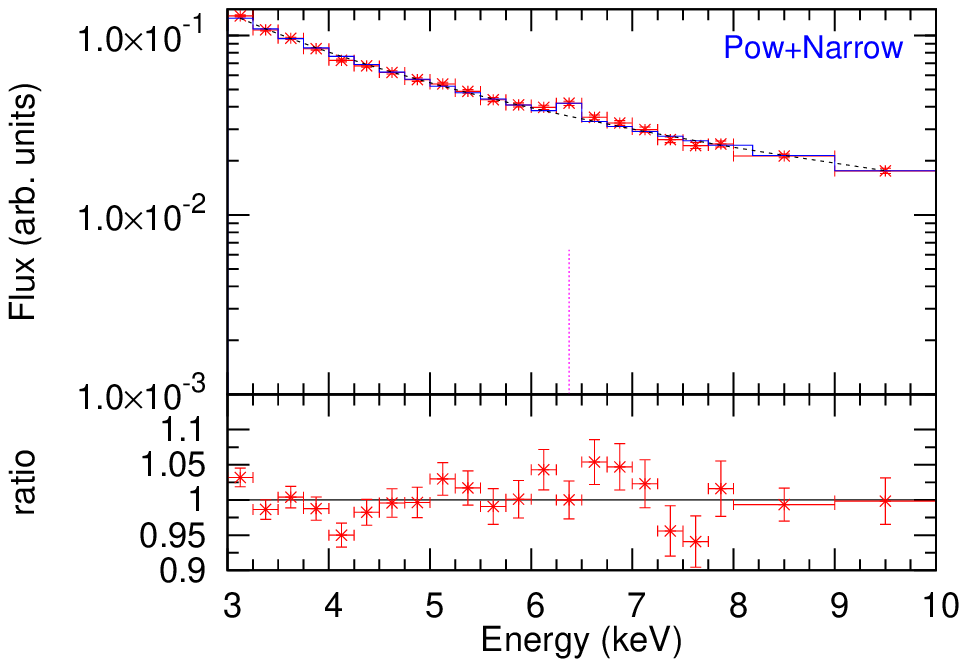} &
\includegraphics[height=5cm]{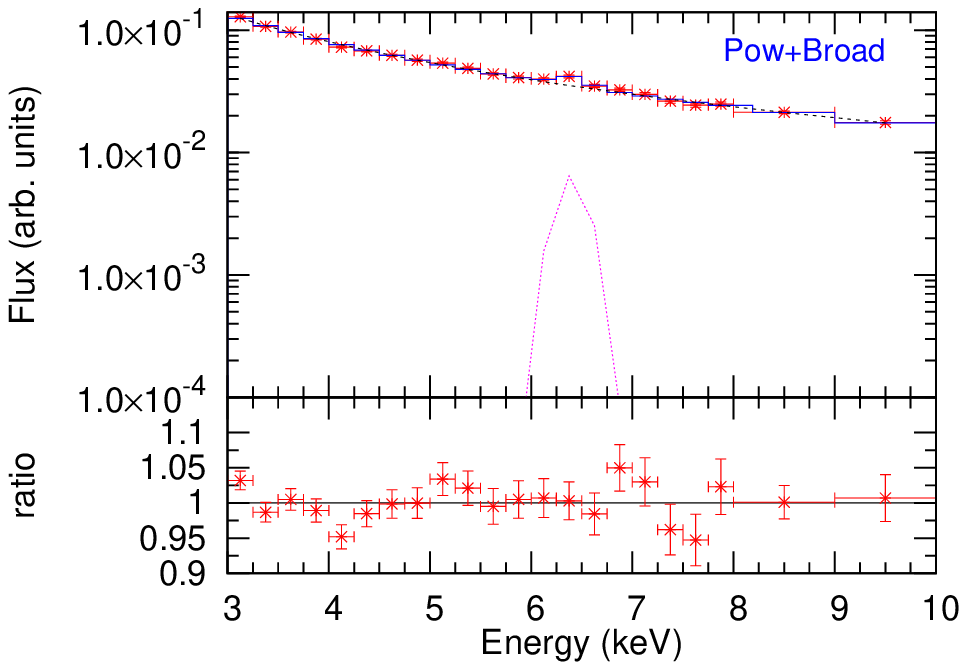} \\
\includegraphics[height=5cm]{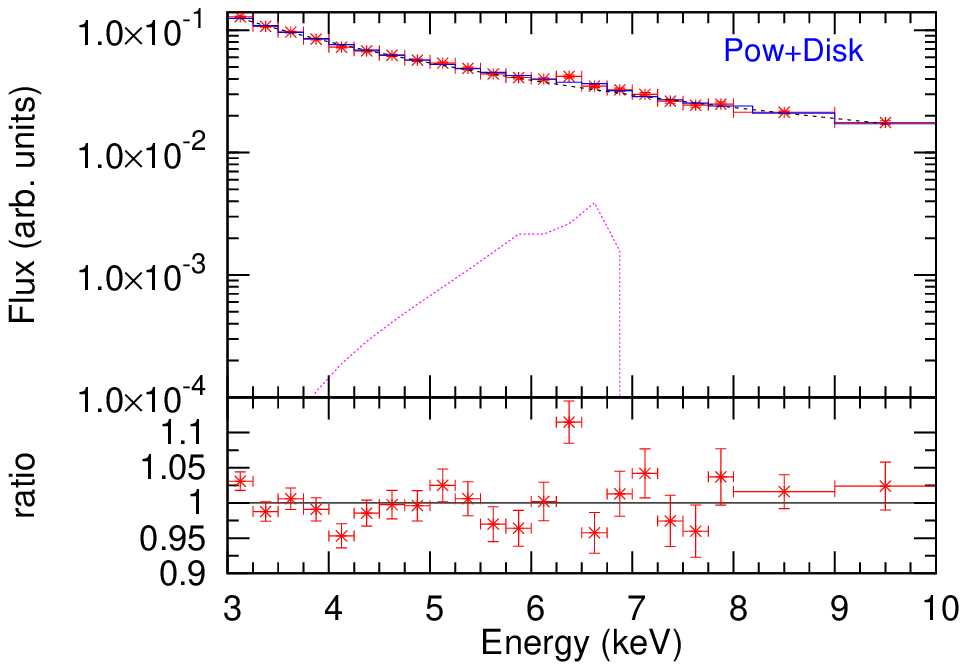} &
\includegraphics[height=5cm]{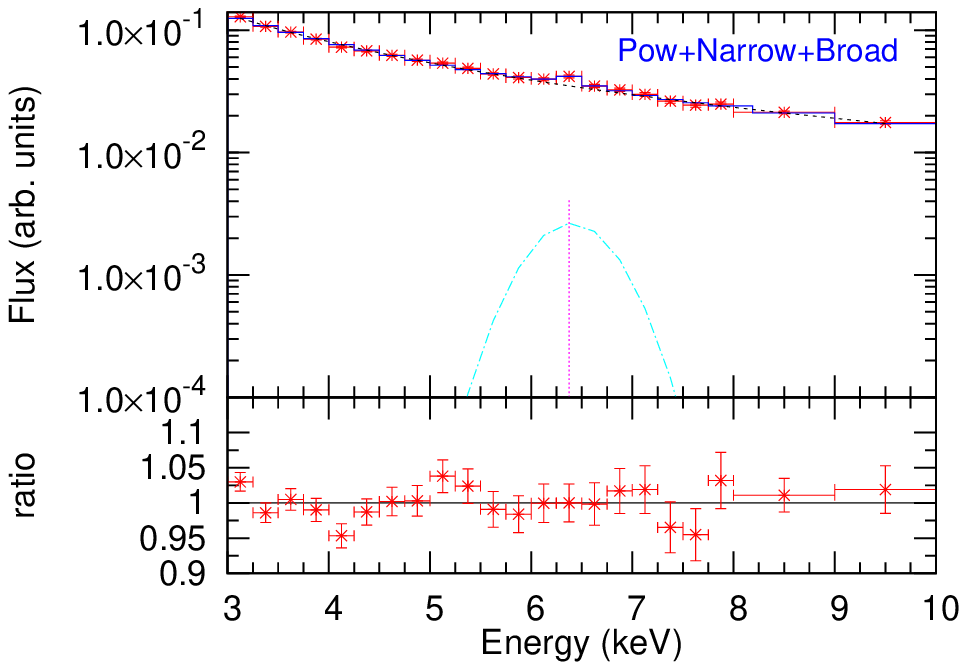} \\
\includegraphics[height=5cm]{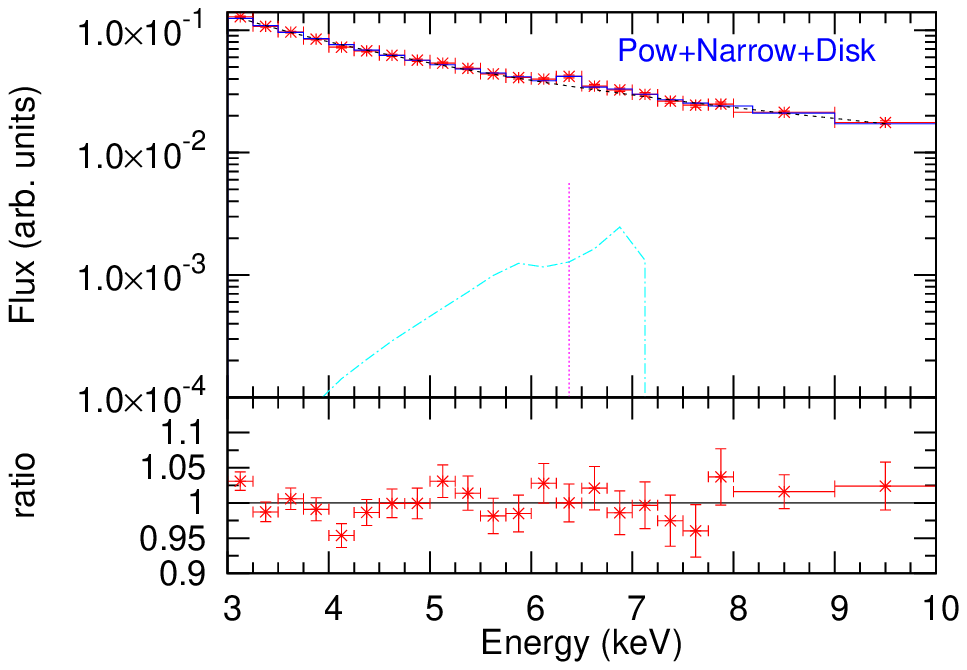} &
\includegraphics[height=5cm]{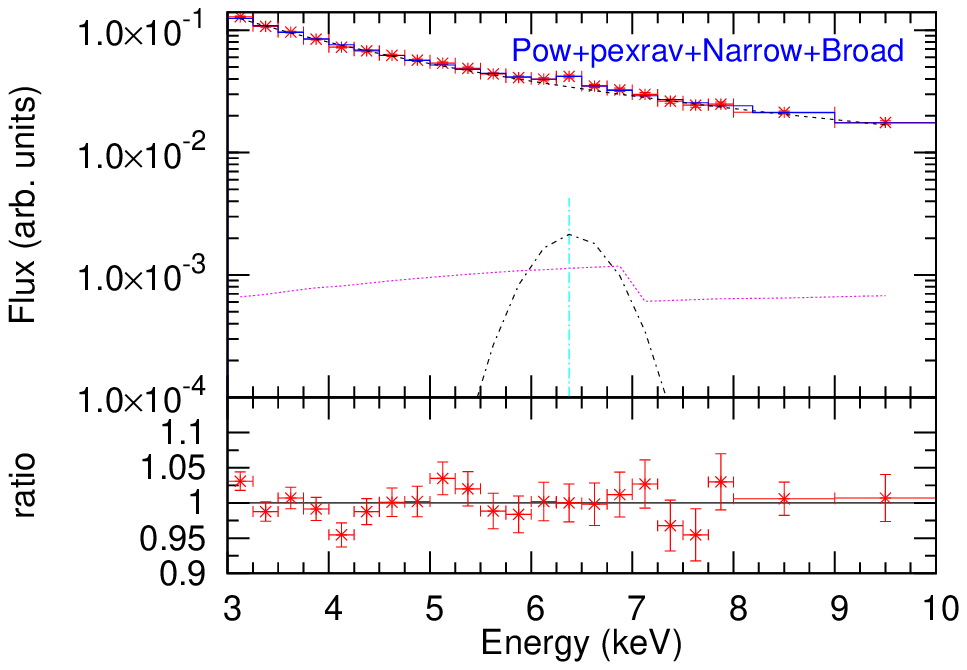}\\
\includegraphics[height=5cm]{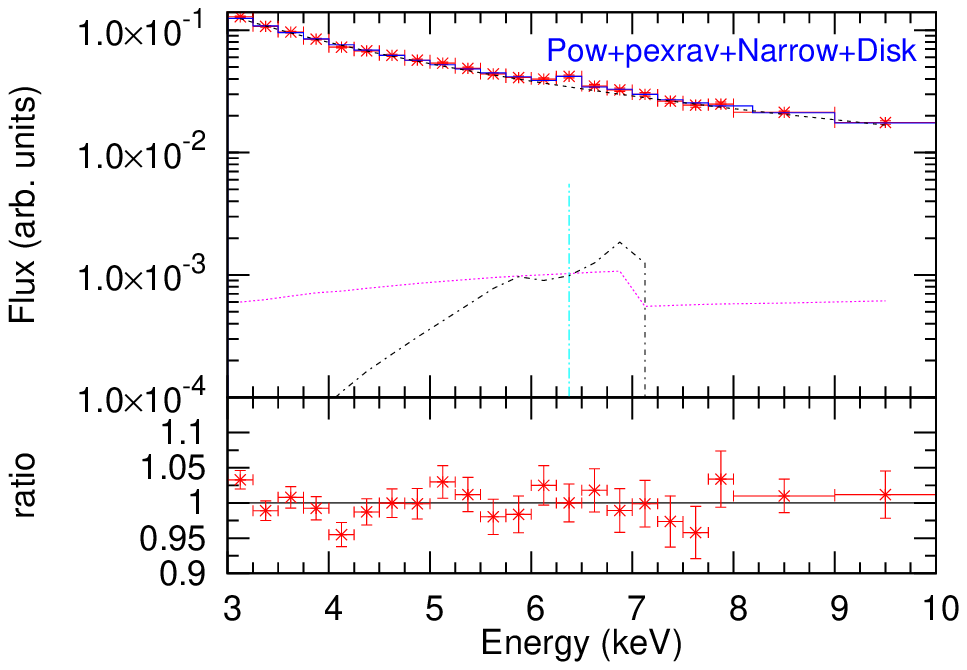}&
\end{tabular} 
\caption{{\it Top panels} in each figure show spectral fit to the {\bfseries ``averaged ratio flux spectrum''} using the quoted model. In the {\it bottom panels} ratio with respect to the quoted model is shown. The model components are also displayed.}
\end{figure*}

\begin{figure*}
\centering
\begin{tabular}{cc}
\includegraphics[height=5cm]{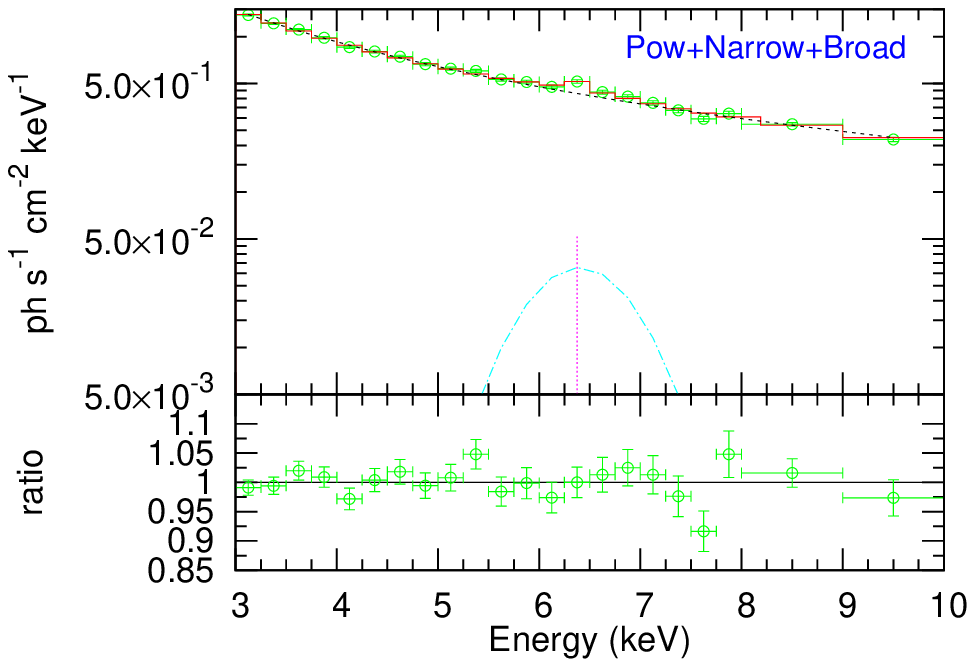} &
\includegraphics[height=5cm]{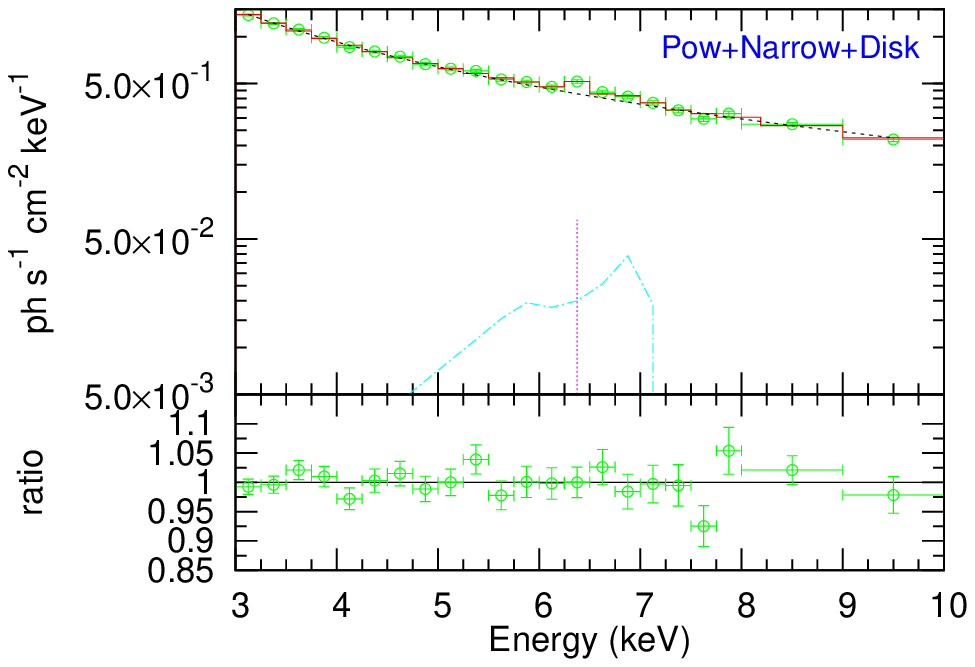} \\
\includegraphics[height=5cm]{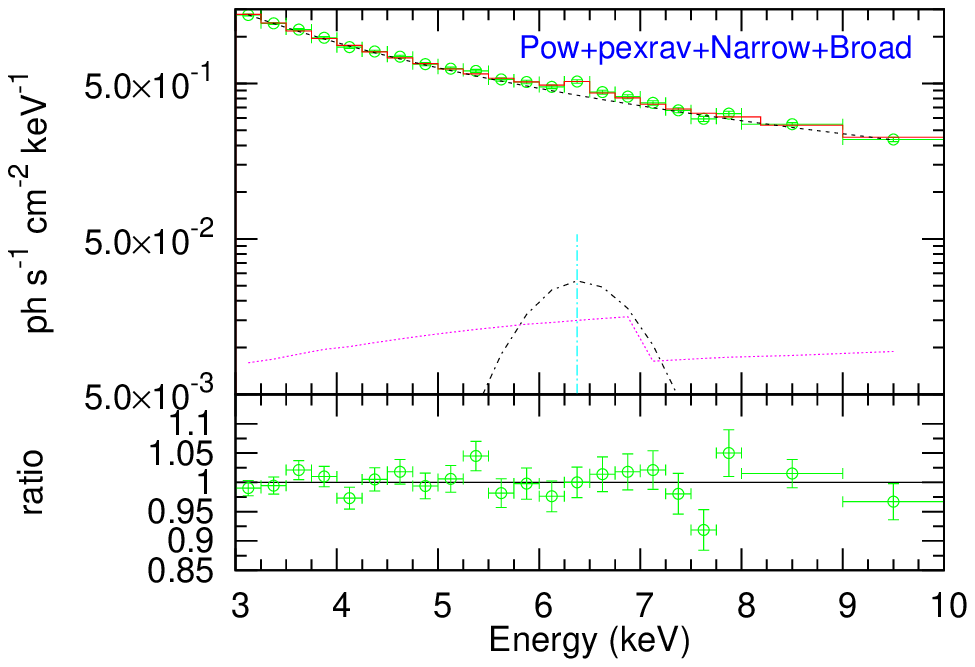}&
\includegraphics[height=5cm]{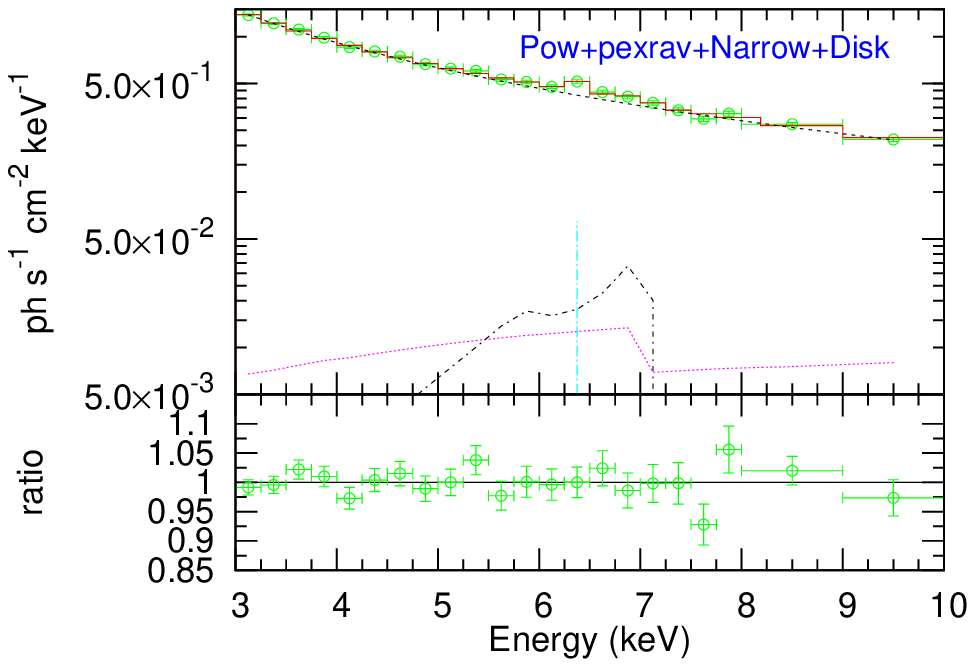} \\
\end{tabular} 
\caption{{\it Top panels} in each figure show spectral fit to the {\bfseries ``averaged X-ray spectrum''} using the quoted model. In the {\it bottom panels} ratio with respect to the quoted model is shown. The model components are also displayed.}
\end{figure*}
\end{appendix}

\end{document}